\documentclass[twocolumn]{revtex4-2}

\usepackage{amsmath,amssymb,amsfonts}
\usepackage{graphicx,subfigure}
\graphicspath{{Figures/}}
\usepackage[x11names]{xcolor}
\usepackage[colorlinks=true,
            linkcolor=Blue3,
            urlcolor=Blue4,
            citecolor=Blue2]{hyperref}
\usepackage{xspace}
\usepackage{cleveref}
\usepackage[utf8]{inputenc}
\usepackage[english]{babel}
\usepackage{natbib}
\usepackage{booktabs}
\usepackage{bm}
\usepackage{txfonts}

\newcommand{\refeq}[1]{\cref{eq:#1}}
\newcommand{\hosp}{\textsl{hosp}\xspace}
\newcommand{\conf}{\textsl{conf}\xspace}
\newcommand{\malawi}{\textsl{malawi}\xspace}
\newcommand{\baboons}{\textsl{baboons}\xspace}
\newcommand{\about}{\ensuremath{\simeq}}
\newcommand\mean[1]{\ensuremath{\langle#1\rangle}}
\newcommand\Ncon{\ensuremath{N_{int}}}
\newcommand\Nint{\ensuremath{N_{int}}}
\newcommand\Nclus{\ensuremath{N_{clus}}\xspace}
\newcommand{\pte}{p.t.e\xspace}
\newcommand\sect[1]{Sect. \ref{sec:#1}}
\newcommand\Fig[1]{Figure \ref{fig:#1}}
\newcommand\Table[1]{Table \ref{tab:#1}}
\newcommand{\SI}{\textit{SI}}
\newcommand{\Sup}[1]{\SI \textit{Appendix, #1}\xspace}
\newcommand{\safeurl}[1]{\href{#1}{#1}}

\usepackage{colortbl}
\definecolor{lightgrey}{rgb}{0.95,0.95,0.95} 
\definecolor{rose}{rgb}{1.,0.75,0.8} 

\begin{document}

\title{How long do we stand our colleagues? \\A universal behavior in face-to-face relations}

% Use letters for affiliations, numbers to show equal authorship (if applicable) and to indicate the corresponding author
\author{St\'ephane Plaszczynski}
\email{stephane.plaszczynski@ijclab.in2p3.fr}
\author{Gilberto Nakamura} 
\affiliation{Universit\'e Paris-Saclay, CNRS/IN2P3, IJCLab, 91405 Orsay, France}
\affiliation{Center for Interdisciplinary Research in Biology (CIRB), Coll\`ege de France, CNRS, INSERM, Universit\'e PSL, Paris, France.}
\date{\today}

\begin{abstract}
\paragraph*{Abstract}
We compare face-to-face interaction data recorded by wearable sensors
in various sociological environments. 
The interactions among individuals display a clear
environment-dependent diversity in agreement with previous analyses.
The contact durations follow 
heavy-tailed distributions although not exactly of power-law type as
previously suggested.
Guided by the common patterns
observed for each relation, we introduce a
variable named the duration contrast, which reveals a common
behavior among all datasets. This suggests that our
tendency to spend more or less time than usual with a given
individual in a face-to-face relation is not governed by social rules
but by a common human trait. Additional data shows that it is the same 
for baboons. Furthermore, we propose a new kind of model to
describe the contacts in a given relation based on the recently
introduced concept of Levy Geometric Graphs. It reproduces the data at 
an impressive level.
The associated Levy index is found to be $\alpha = 1.1$ on all the
datasets, suggesting a universal law for primates and opening
many exciting perspectives.
\end{abstract}

\maketitle

\begin{center}
\colorbox{lightgrey}{
\parbox{.42\textwidth}{%
  \vskip10pt
  \leftskip10pt\rightskip10pt
  \footnotesize
  \paragraph*{Significance statement}
By comparing face-to-face interaction data obtained in very different
environments (a hospital in France, an international
conference in Italy, a small village in Africa)
we show that the deviation from the average duration of contact
between each pair of individuals (the duration contrast) follows
in each case a very similar statistical law. Surprisingly, the same distribution is
observed among a population of baboons. It suggests that our tendency
to deviate from the usual (mean) time spent in any face-to-face
relation is primarily independent of the sociological environment and governed by its own rules.
To explicit them, we propose an original model based on a random
Levy-walk in a 2D space of relation, that perfectly reproduces the data.
  \vskip10pt
 }
}
\end{center}

\section*{Introduction}
Since the advent of the Internet at the dawn of the XXst 
century, the quantity of digital data describing our behavior
has inflated, offering to scientists an unprecedented
opportunity to study human interactions in a more 
quantitative way. This 
 opened the field of sociology to data-analysis and 
from the hard-science community, came the tacit idea that 
complex human behaviors can be modeled as simple systems 
\cite{Song:2010,Stehle:2010,Zhao:2011,Starnini:2013,Sekara:2016,Flores:2018}. 
With the rapid development of mobile technologies (GPS, Bluetooth,
cellphones) a lot of effort was first put in trying to
capture the patterns of human mobility (for a review, see \cite{Barbosa:2018}).
A more local picture of our everyday social interactions can be obtained
using dedicated proximity sensors.
Following a pioneering experiment that equipped conference participants
with pocket switched devices \cite{Hui:2005,Scherrer:2008}, 
the \textit{sociopatterns} collaboration
(\safeurl{www.sociopatterns.org}) developed wearable sensors
that allow to register the complex patterns of face-to-face
interactions \cite{Cattuto:2010,Barrat:2014}. The radio-frequency
signal is only recorded if two individual
are in front of each other for a duration of a least 20 s (which is
the timing resolution). We note that, from a sociological point of view,
a distance below 1.5 m covers the traditional 
\textit{private} ($<50$ cm), \textit{personal} ($<$1.2 m) and
\textit{social} ($<3.5$ m) zones. 
The goal is not only to analyze social interactions but also to
understand how information (or even a disease) spreads over a real dynamical
network \cite{Stehle:2011,Isella:2010,Starnini:2012,Genois:2015}.
Those sensors were worn by volunteers in several work-related environments: 
scientific conferences \cite{Cattuto:2010,Isella:2010,Stehle:2011},
a hospital ward \cite{Vanhems:2013}, an office \cite{Genois:2015} and at
school \cite{Fournet:2014,Mastrandrea:2015}. 
As part of a UNICEF program, they were also used to
characterize social exchanges in small villages in Kenya and Malawi \cite{Kiti:2016,Ozella:2021}
and for ethological studies on baboons \cite{Gelardi:2020}.
It was rapidly noticed (see \cite{Barrat:2015} for a short review)
that the duration of face-to-face contacts follows
a broad distribution, sometimes qualified a bit rapidly as a
``power-law'', and that it was similar among participants in several
environments. 

We wish to revisit here in detail this assertion.
Most importantly, the comparisons performed so far applied to some 
very similar sociological environments 
(conferences, high-school, office...) which are typically occidental
(french), educated and with a scientific background. It is  
unclear to which level they are general.
We thus also compare those data to the ones from rural Malawi,
and baboon interactions.

After describing our data selection and some
methodological difference with previous studies in \sect{method}, 
we will show in \sect{struct}, that social interactions between
individuals are very different in each environment.
We will then study the temporal aspects of each relation in
\sect{temporal} introducing the concept of contrast for contact
duration. It will be shown that its distribution is essentially the
same for each dataset and for each relation individually.
We then present in \sect{model} a model completely different
from previously proposed ones.
Indeed, it is not agent-based but describes the relation itself.
Based on a recent theoretical work on Levy Geometric Graphs
\cite{Plaszczynski:2022}, we will show that it reproduces perfectly the data.
This model raises questions that will be discussed in
\sect{discussion}.
Some extra information, referred to in the text, is given in the \textit{Supporting Information} (\SI) document.

\section{Material and methods}
\label{sec:method}

\subsection{Datasets}

We have chosen 4 datasets sociologically most dissimilar.

\begin{enumerate}
\item \hosp : these are early data collected over 3 days \footnote{here
    and in the following, we will only consider complete (24 h) day periods.} on 75
  participants in the geriatric unit of a
  hospital in Lyon (France) \cite{Vanhems:2013}. Most
  interactions (75\%) involve nurses and patients (not doctors) so 
  we consider that sample to be sociologically quite different from
  the next (scientific) one.
\item \conf: these are also some early classical data from the ACM Hypertext 2009 
  conference (\safeurl{www.ht2009.org}) that involved about a hundred of participants
  for 3 days \cite{Isella:2010} in Torino (Italy). The audience is
  international with a scientific background.
  There exist also some data taken at another conference in Nice in 2009 (SFHH, \cite{Genois:2018})
  with more participants, but we prefer to use the former which has a
  number of individuals comparable to the other datasets. However we
  have checked that we obtain similar results with the SFHH data.
\item \malawi: those proximity data were taken in a small village of the district
  of Dowa in Malawi (Africa) where 86 participants agreed to participate for
  13 (complete) days. Interestingly those data contain both extra 
  and intra-household interactions, although we will
  not split them in the following. This community consists
  essentially of farmers. 
\item \baboons Those data were taken at a CNRS Primate Center near Marseille
  (France) where 13 baboons were equipped with the sensors for a
  duration of 26 days. The goal was to study their interactions, and
  study how conclusions reached from data-analysis match those
  provided by standard observation.
\end{enumerate}

With that choice, we span very different sociological environments.
We have also analyzed a few other datasets collected at the SFHH conference, an
office and a high-school. They give similar results 
but we consider them as sociologically closer to the \conf one.
A valuable interest, is that those data were taken with the very same
devices, and can be compared immediately, minimizing possible sources of systematic errors.

\subsection{Differences with previous studies}

Previous studies considered the overall temporal properties of
interactions, i.e. without differentiating the pair of people interacting.
In this work we will put accent on the temporal properties of each
pair separately.

Probability distribution functions (p.d.f) are often estimated by
histograms, i.e. by counting the number of samples that falls within
some bins. But for heavy-tailed distributions the size of the bins is
delicate to choose. 
With a constant size binning, several bins end up empty for large
values. Using a logarithmically increasing binning is neither a
solution since it supposes that the
distribution is constant on the wide range of last bins.
Following \cite{Newman:2005}, we will use instead the \textit{probability to
  exceed} function (\pte, also known as the ``complementary cumulative
distribution function'', CCDF) which is computed simply by sorting the
samples and plotting them with respect to their relative frequency. In
this way, one does not need to define a binning.
If the p.d.f follows a power-law its \pte also.
Conversely, if the \pte plotted on logarithmic axes is not linear, its
p.d.f is not a power-law neither.

%graphs
\section{Social interactions between individuals}
\label{sec:struct}

\begin{figure}[htbp]
  \centering
  \subfigure[\hosp]{\includegraphics[width=0.45\linewidth]{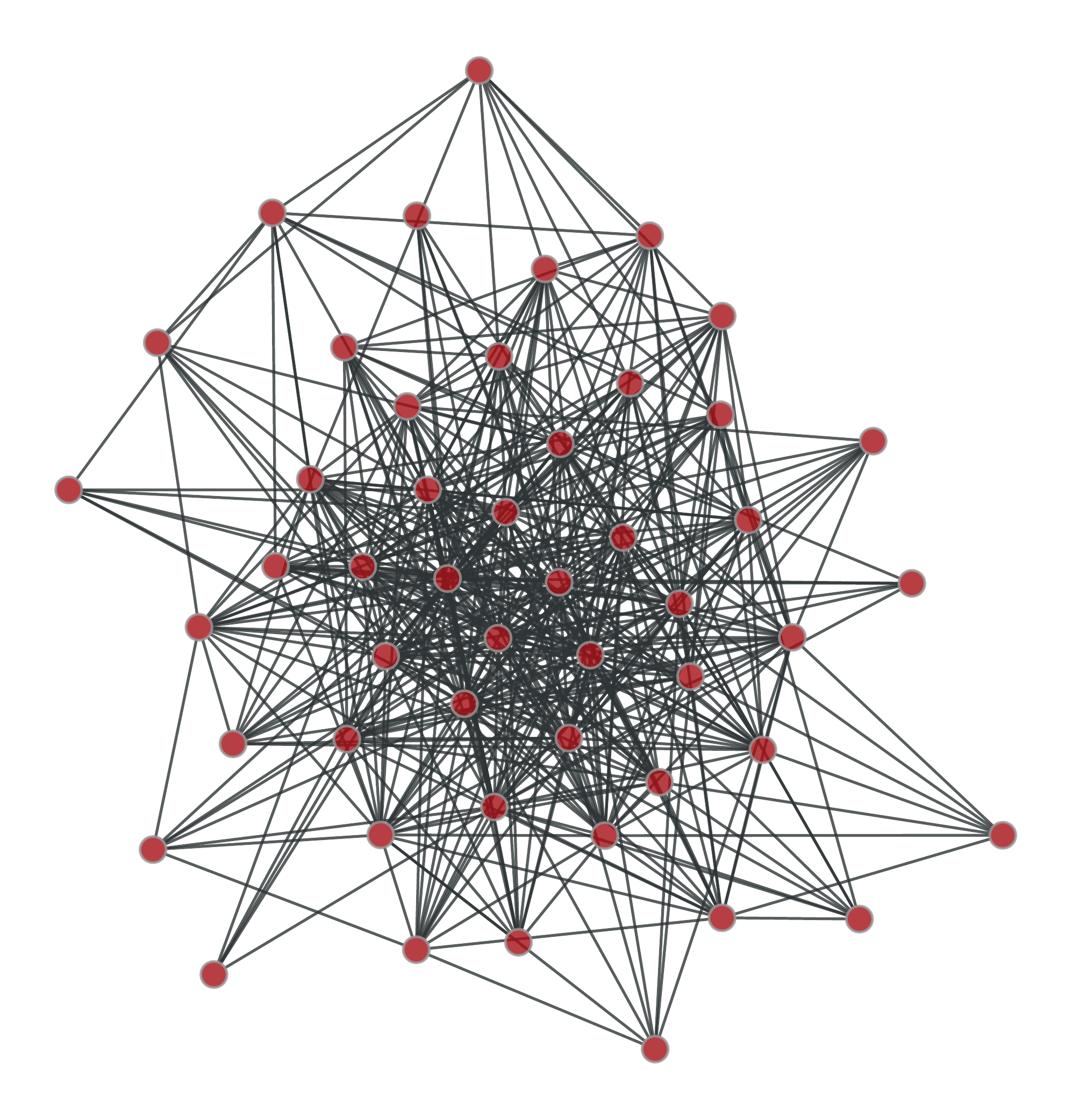}}
  \subfigure[\conf]{\includegraphics[width=0.45\linewidth]{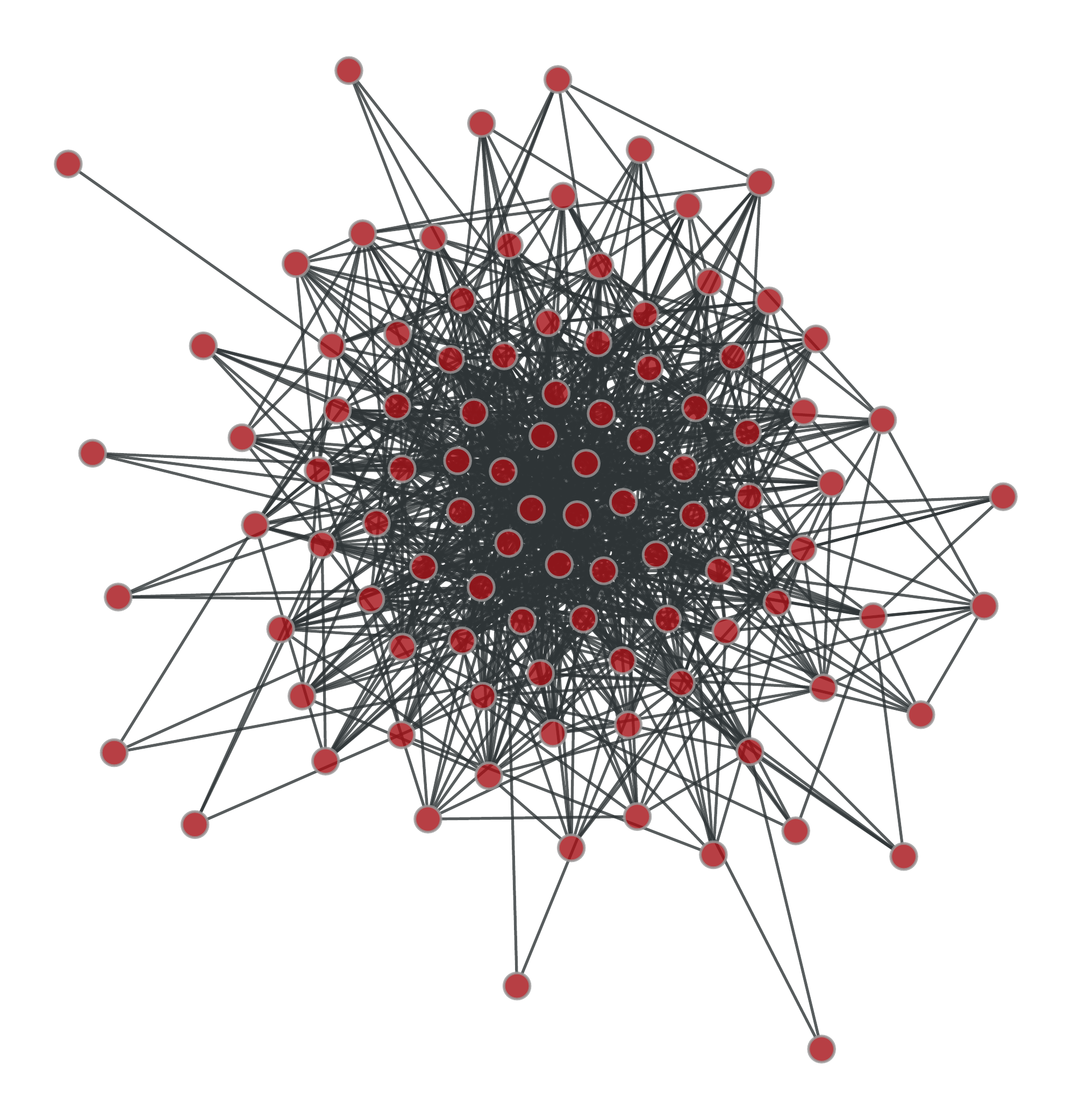}}
  \subfigure[\malawi]{\includegraphics[width=0.45\linewidth]{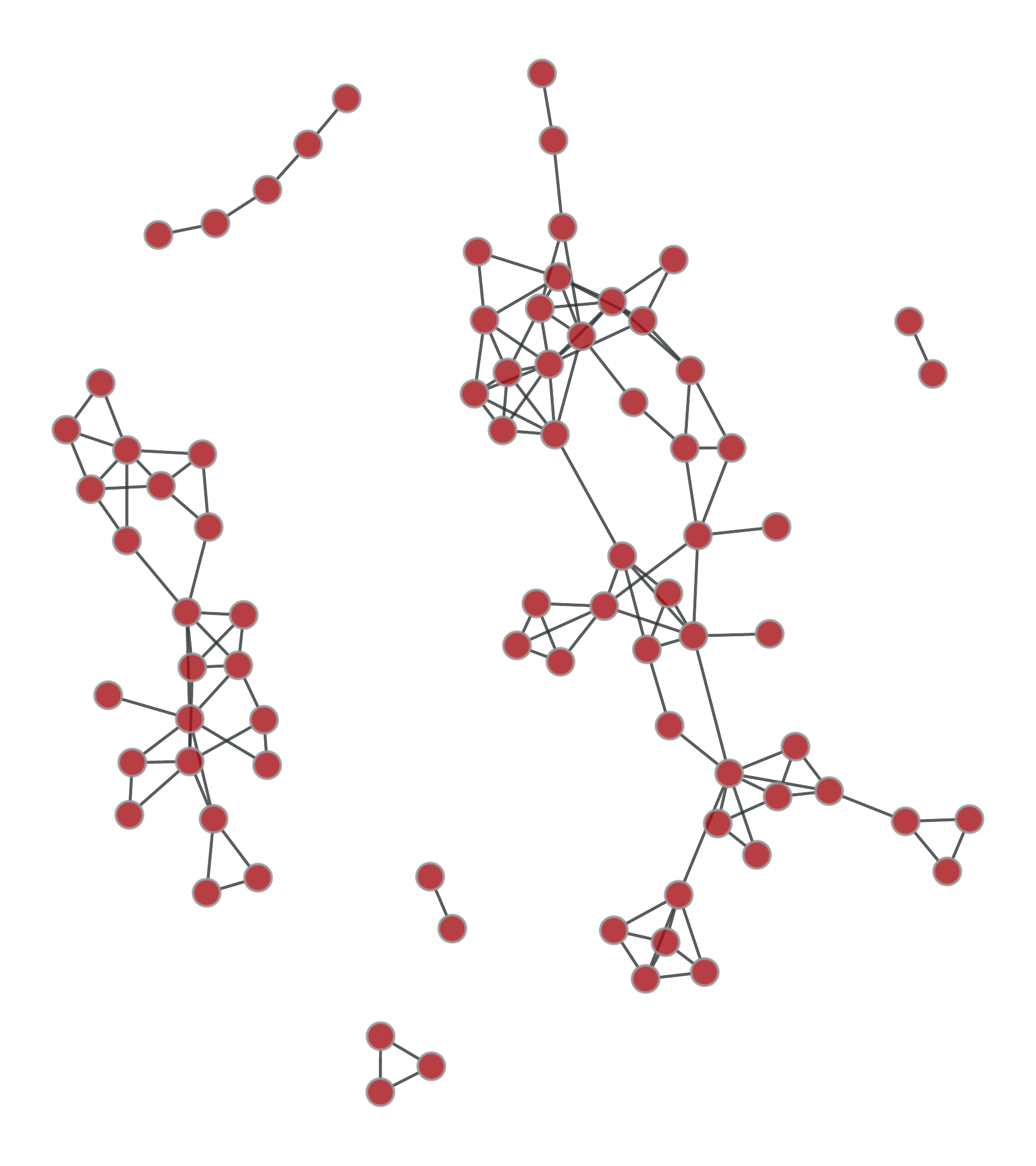}}
  \subfigure[\baboons]{\includegraphics[width=0.45\linewidth]{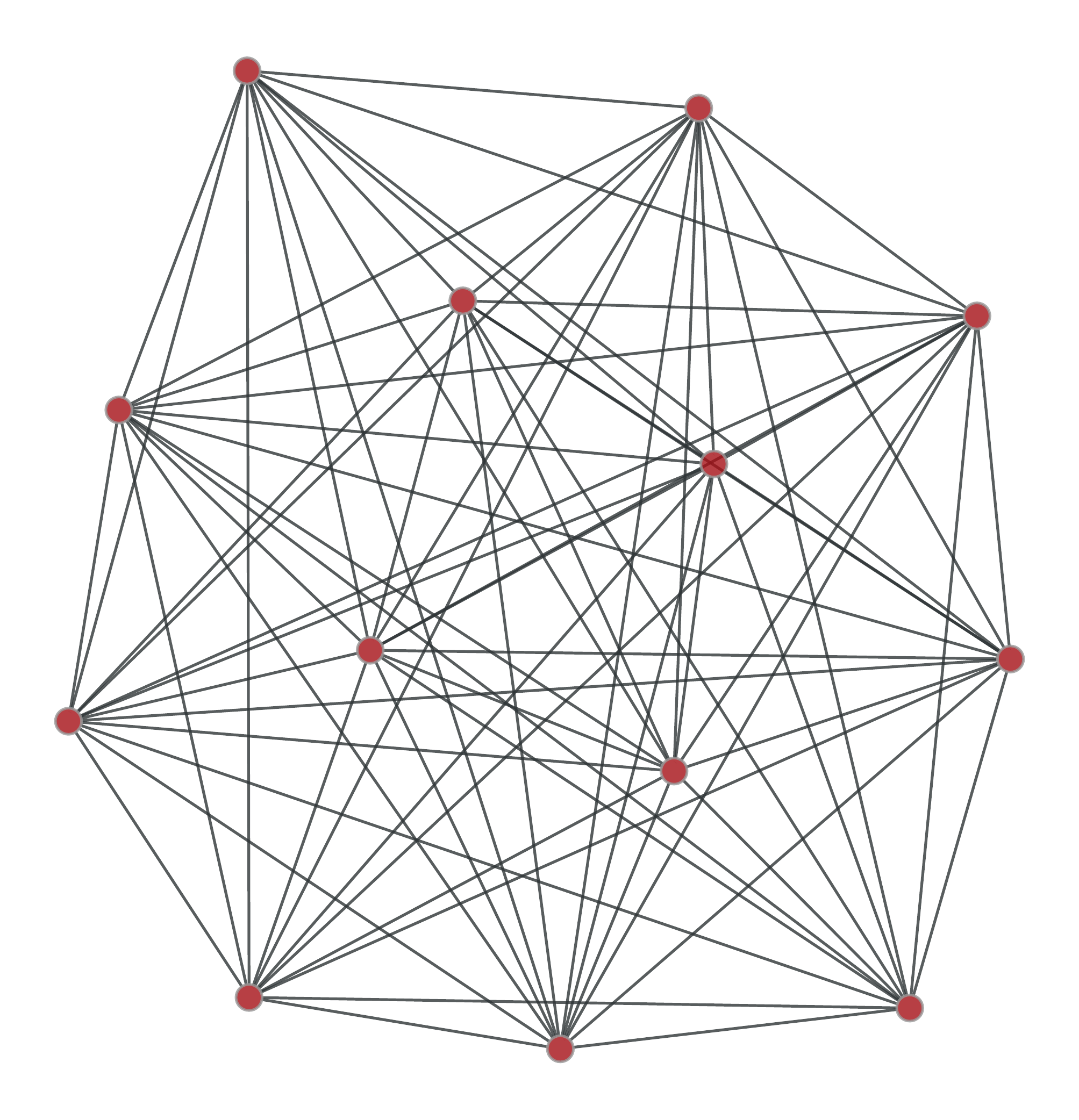}}
\caption{Aggregated graphs of interactions over one day for our 4
  datasets. Vertices (red points)
  represent agents and there is a link (edge) if a face-to-face interaction
  occurred for more than 20 s.}
\label{fig:graphs}
\end{figure}

The structure of social interactions between individuals 
can be studied through time-aggregated graphs 
representing any ($>$ 20 s) face-to-face interaction in a given
time window. Since those interactions most often repeat on a day-by-day
basis, we give results on a complete day (24 h).
The graphs obtained for the first day are shown on \Fig{graphs}, and we checked that
they are similar for the others.

The graphs for the \hosp and especially the \conf datasets show a strongly connected
core. The \malawi one is much sparser, while the \baboons one is
almost complete showing that each animal interact with all the others.

\begin{table}
  \centering
  \caption{\label{tab:graphs} Properties of time
    aggregated graphs on each dataset per day. Uncertainties 
    are the standard deviations between the days. 
    $T$ is the number of (complete) days in the dataset. $N$ the number
    of interacting agents.
    $\mean{k}$ is the mean degree, i.e. the average number of agents each
    individual interacts with during one day. 
    $\mean{w}$ is the mean weight where the weights
    specify the total duration of a single relation \cite{Barrat:2004}.  
    Mean strength \mean{s}
    which represents the average total interaction time per individual.
  }
  \begin{tabular}{lrrrrrrr}
    group & T &  N & \mean{k} & \mean{w}(mins) & \mean{s} (mins) \\
    \midrule      

    \hosp & $3$ & $ 49\pm1 $ & $ 18\pm1 $ & $ 6\pm13 $ & $97\pm101$\\

    \conf & $3$ & $ 100\pm3 $ &  $ 20\pm1 $ & $ 2\pm14 $ & $ 46\pm63$\\

    \malawi & $12$ & $ 70\pm4 $ &  $ 3\pm1 $ &$24\pm37$ & $65\pm74$\\

    \baboons & $26$ & $ 13\pm1 $  & $ 11\pm1 $ & $8\pm11$ &$87\pm53$\\
\bottomrule
    \end{tabular}
  \end{table}

\Table{graphs} gives a more quantitative view of some of the graph's properties.
The number of different people met per day (the degree of the graphs) is
about 20 in both the hospital and the conference environments.
As is apparent in \Fig{graphs}(c), it is is much smaller in the rural
community (3). 
But the interaction times are longer (\about 25 min) which reflect
different sector of activities (agrarian and including inter-housing
relations for the \malawi data). 

The strength of the relation represents the total time per
individual spent interacting
with others per day. It is essentially the product of the mean number
of people met per day and the
time spent interacting with them ($\mean{s}\simeq\mean{k}\mean{w}$).
It varies
by a factor of two (from 45 min to 1.5 h) although the large
standard-deviations indicates important daily variations due to the 
heavy-tail of the distribution.

The comparison to the \baboons dataset should be handled with care
since there is a much smaller number of agents (13).
Since the baboons interact essentially all with each other
(\Fig{graphs}(d)), their mean degree is bounded, $\mean{k}\simeq N$.
On the other hand, their small number probably increases their
interaction duration  
($\mean{w}$) so that the strength of their relation is finally similar to
that of the human groups.

The goal of this short section is not to enter into the detail of the
differences but to
highlight that, as expected, those heterogeneous sociological groups
have some very
different interaction patterns between individuals.

\section{Face-to-face temporal relation}
\label{sec:temporal}

\begin{figure}[ht!]
  \centering
  \includegraphics[width=.9\linewidth]{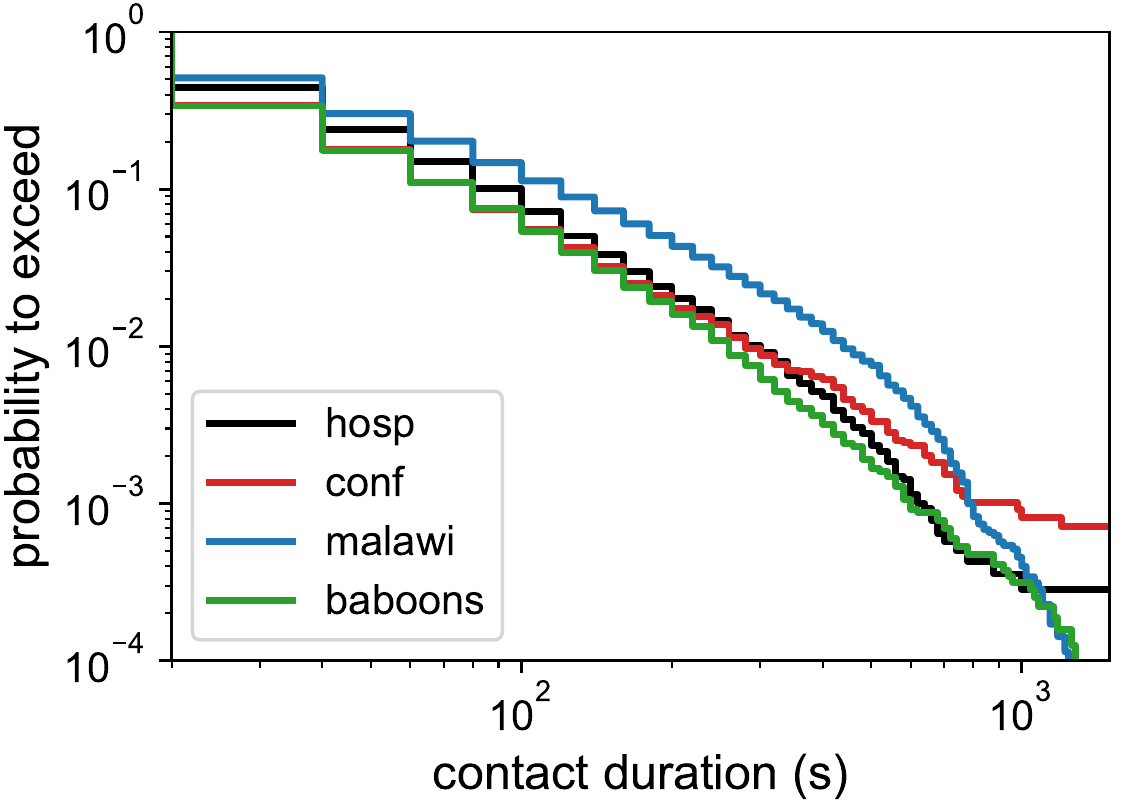}
  \caption{\label{fig:tij} Distribution (\pte) of the contact
    duration on the four datasets (all days used).}
\end{figure}

We consider now the duration of all interactions between individuals. \Fig{tij} shows the
distribution for each group. They are indeed ``heavy-tailed''; most
interactions are short time (at the minute level) but some may drift
up to an hour.
Although it is often claimed that
these distributions follow a ``power-law'' (which would look linear in the
logarithmic representation)
this more precise approach, using the \pte, shows that it is only true
over a limited range or rather that a curvature exists. 
Interactions for people in \malawi tend to last longer than for
all the others.
The baboons' duration of interaction
is similar to the human ones (as noticed in \cite{Gelardi:2020}), although there
are some sizable differences at short times, somewhat squeezed by the
logarithmic scale.
Overall, at the level of precision we wish to investigate, we
consider these distributions as quite different and not following a power-law.

We focus on the detail of each relation, i.e. pair of individuals,
corresponding to edges in time-aggregated graphs.
Each one consists in a set of intervals measuring the beginning and
end times of each interaction at the resolution of the instruments (20 s). 
There is a varying number of interactions (intervals) per relation, 
that we call $\Ncon(r)$.
In the following we will only consider the duration of the
interactions (postponing inter-contact durations to the Discussion)
that we note $\{t_i(r)\}$. They are variable-size timelines. Unless
otherwise specified, they are expressed as a number of resolution steps.

\begin{figure}[ht!]
  \centering
  \subfigure[]{\includegraphics[width=.9\linewidth]{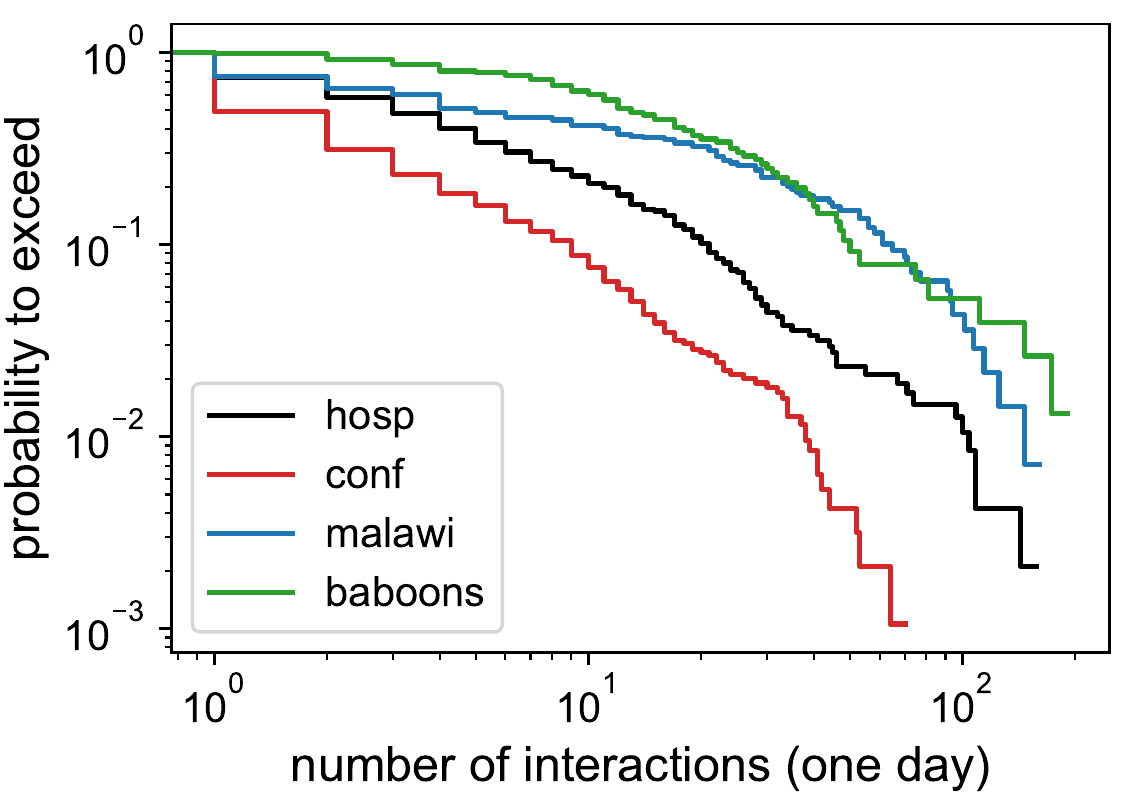}}
  \subfigure[]{\includegraphics[width=.9\linewidth]{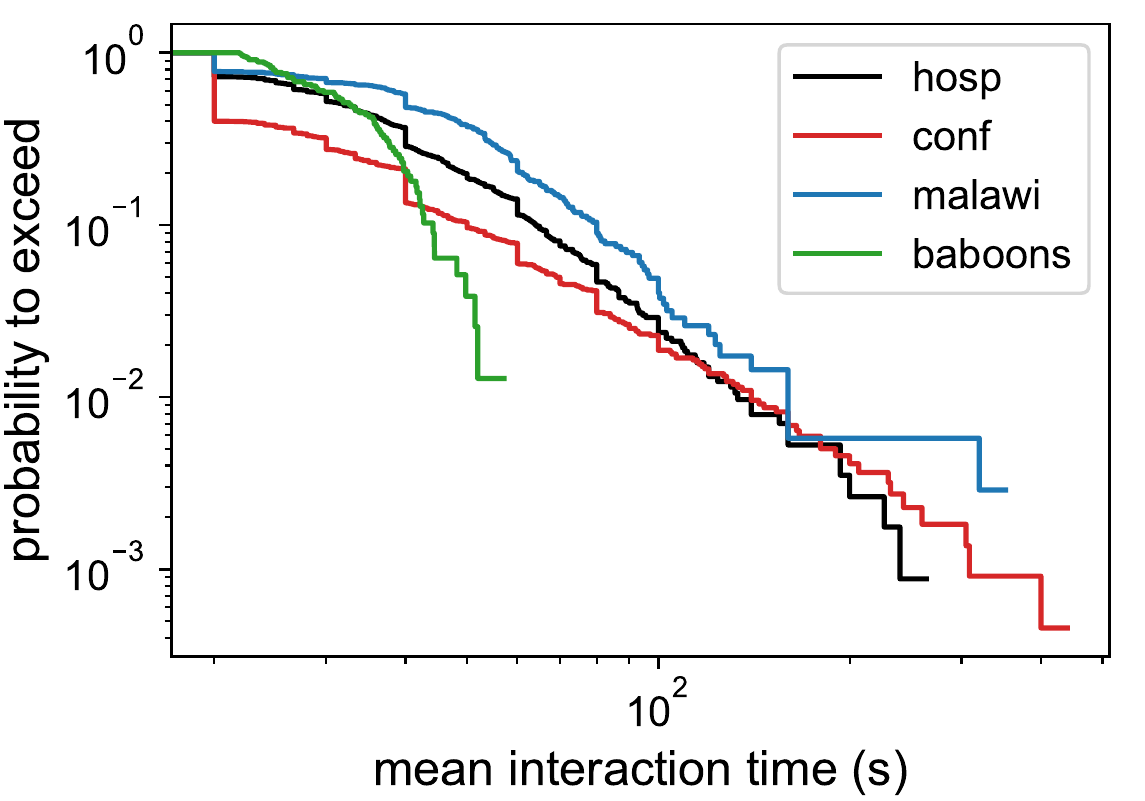}}
  \caption{\label{fig:Ncon} General characteristics of temporal
    relations on the 4 datasets. (a) Distribution (\pte) of the number of
    interactions per relation for one day, and (b) of the mean interaction
    time. To gain precision, we use the complete datasets for the latter.}
\end{figure}

The number of registered interactions for a given pair clearly depends on the
total duration of the experiments (Table \ref{tab:graphs}) but we may
compare it for one day, which then corresponds to a rate of
interactions per day.
The distribution of this variable is shown in \Fig{Ncon}(a). 
It is clearly different for each group. 
People at the conference tend to interact (with the same person) less
often. Frequently (65\%) it is only once per day, against 25\% for the \hosp and
\malawi datasets, and 3\% for \baboons.

The mean interaction time per relation
\begin{align}
\label{eq:meantr}
  \bar t(r)=\dfrac{1}{\Ncon(r)}\displaystyle{\sum_{i=1}^{\Ncon(r)} t_i(r)},
\end{align}
is shown in \Fig{Ncon}(b). 
Here again distributions are heavy-tailed and different. There is a
marked difference between animals and humans, the former interacting
for shorter times.

We now introduce the contact duration contrast of a relation
(or simply ``contrast'')
as the duration of each interaction scaled by its mean value
\begin{align}
  \delta_i(r)=\dfrac{t_i(r)}{\bar t(r)}.
\end{align}
It describes deviations with respect to the mean time of the relation
and is a dimensionless quantity. We emphasize that the scaling by
$\bar t(r)$ is different for each pair of individuals (relation).

\begin{figure}[ht!]
  \centering
  \subfigure[]{\includegraphics[width=.8\linewidth]{cdc_all_log}}
  \subfigure[]{\includegraphics[width=.8\linewidth]{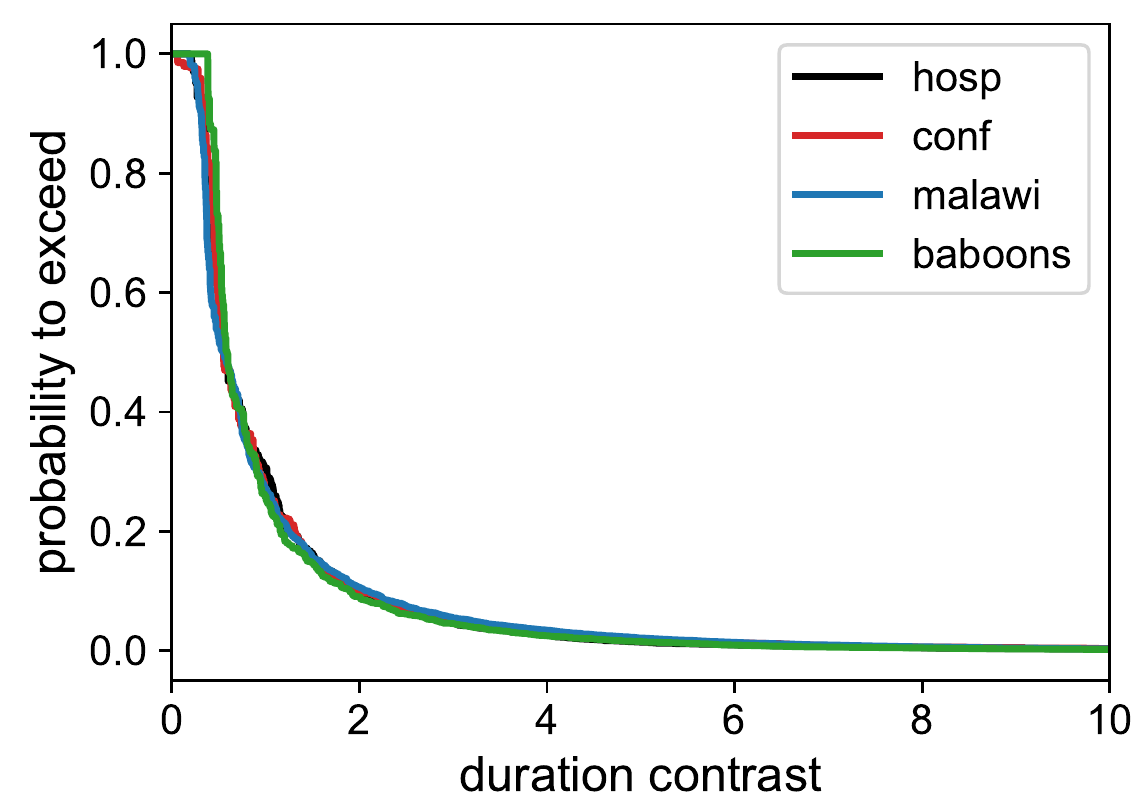}}
  \caption{\label{fig:cdc_all} Distributions (\pte) of the duration
    contrast obtained for all relations  within the same group 
    satisfying $\Ncon(r)>50$ in logarithmic (a) and linear scales (b).
    Due to
    shorter data taking periods (see \Table{graphs}),
    distributions for the \hosp and \conf datasets starts to become
    noisy for contrast above $\simeq$10.}
\end{figure}

When there are few samples, the arithmetic mean (\refeq{meantr}) is a 
poor estimate of the true mean-time value. Furthermore,  
it is strongly correlated to the individual samples. The measured contrast is
then very noisy.
We thus apply a cut to select only timelines with a large
number of samples. Since the time distributions are heavy-tailed, 
we require at least 50 samples to correctly estimate the
mean ($\Ncon(r)>50$). 
Using the full datasets, we are left with respectively 
57, 26, 91 and 70 timelines for the \hosp, \conf, \malawi and \baboons
datasets.
We show the \pte distributions of the contact duration contrast for
the 4 groups in \Fig{cdc_all}.
 
They are strikingly similar up to a contrast of about 10.
Small differences, barely noticeable in the linear representation
\Fig{cdc_all} (b), happen at low scales, but as we shall explain in
\sect{model}, they are most probably due to the finite resolution of the instrument.

If we compare this distribution to \Fig{tij}, we see that by dividing
the duration by each individual mean-time, we have standardized the
results. Strictly speaking,
\Fig{cdc_all} includes the $\Ncon>50$ cut on the number of interactions
while \Fig{tij} does not, but a similar dispersion is observed on the latter
when we include the cut (\Sup{S1}).  
Similar results are obtained at another conference (SFHH), in an office
and in a high-school (\Sup{S2}).

\begin{figure}[ht!]
  \centering
  \includegraphics[width=.8\linewidth]{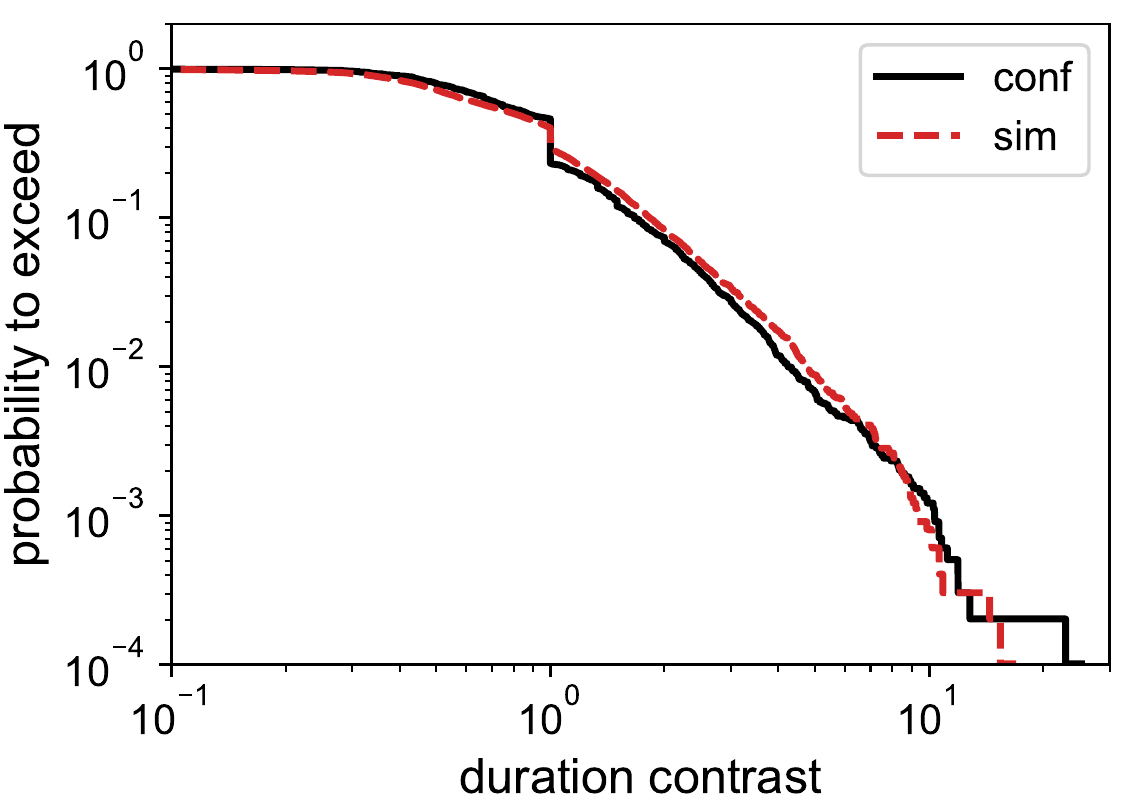}
  \caption{\label{fig:conf_sim} Distributions (\pte) of the duration
    contrast obtained for all relations in the
  \conf dataset and simulations produced using the corresponding
  \Fig{cdc_all} distribution (see text for details). The dip at 1
  comes from numerous cases (65\%) where \Ncon=1 always leads to $\delta=1$.}

\end{figure}

One may think that this ``universal'' distribution is only valid for
high-rate interactions since we have used the $\Ncon(r)>50$ cut.
However this is only an experimental
limitation; with a longer data-taking period most timelines would
exceed 50 interactions and the mean time values would be known precisely.
We argue that the universality is a more general feature and that the
cut, used to clean the data, did not alter the underlying process but
does only allow to reveal it.
First we note that similar results are
obtained with a lower cut value $\Ncon>30$ (\Sup{S3}). 
We then show that we can still reproduce the contrast distribution 
without any cut, using only the distributions with the cut (\Fig{cdc_all}).
To this purpose we perform Monte-Carlo simulations. For a given
dataset, for each relation (without any cut), we draw $\Ncon(r)$ random numbers 
following \Fig{cdc_all} distribution to obtain
$\delta_{i=1,\cdots,\Ncon}$ contrast values.
Those samples are obtained from the distribution with the
$\Ncon(r)>50$ cut, so with precise mean values that we call $\mu$.
We may mimic the statistical fluctuations due to any $\Ncon(r)$ value,
by using the ratio
 \begin{align}
  \delta_i^{mes}=\dfrac{\delta_i}{\tfrac{1}{\Ncon}\sum_i \delta_i}
   =\dfrac{t_i/\mu}{\tfrac{1}{\Ncon} \sum_i t_i/\mu}
   =\dfrac{t_i}{\bar t}
\end{align}
since $\mu$ actually cancels out. 
We compare the measured contrast distribution to the one observed on data, this
time without any \Ncon(r) cut, in \Fig{conf_sim} for the \conf
dataset. We reproduce correctly the whole contrast distribution
using only the \Fig{cdc_all} one obtained with $\simeq$1\% of the data ($\Ncon>50$).
Similar results are obtained for the other datasets (\Sup{S4.1}).
This shows that the contrast distributions obtained on the large sample
statistics are sufficient to reproduce any number of interactions, including very low ones.
In other words, the $\Ncon(r)>50$ cut only cleans the data 
without affecting the underlying ``true'' contrast distribution.

\begin{figure}[ht!]
  \centering
  \subfigure[\hosp]{\includegraphics[width=0.49\linewidth]{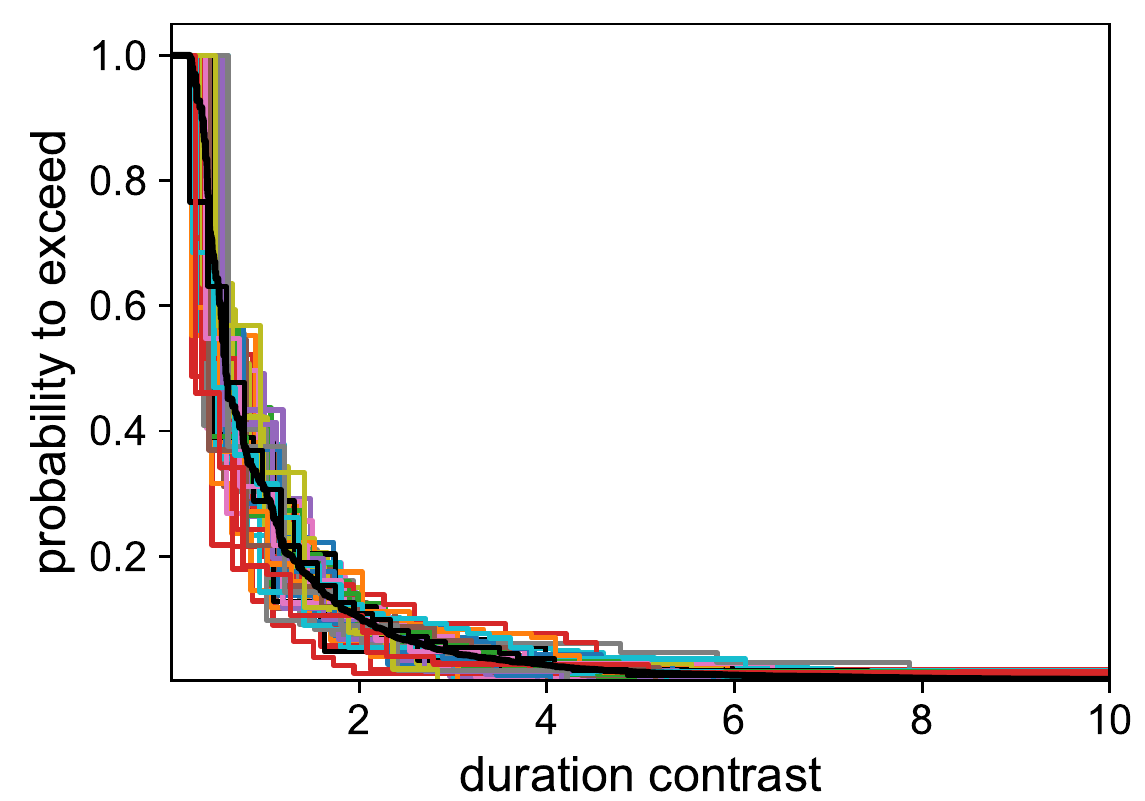}}
  \subfigure[\conf]{\includegraphics[width=0.49\linewidth]{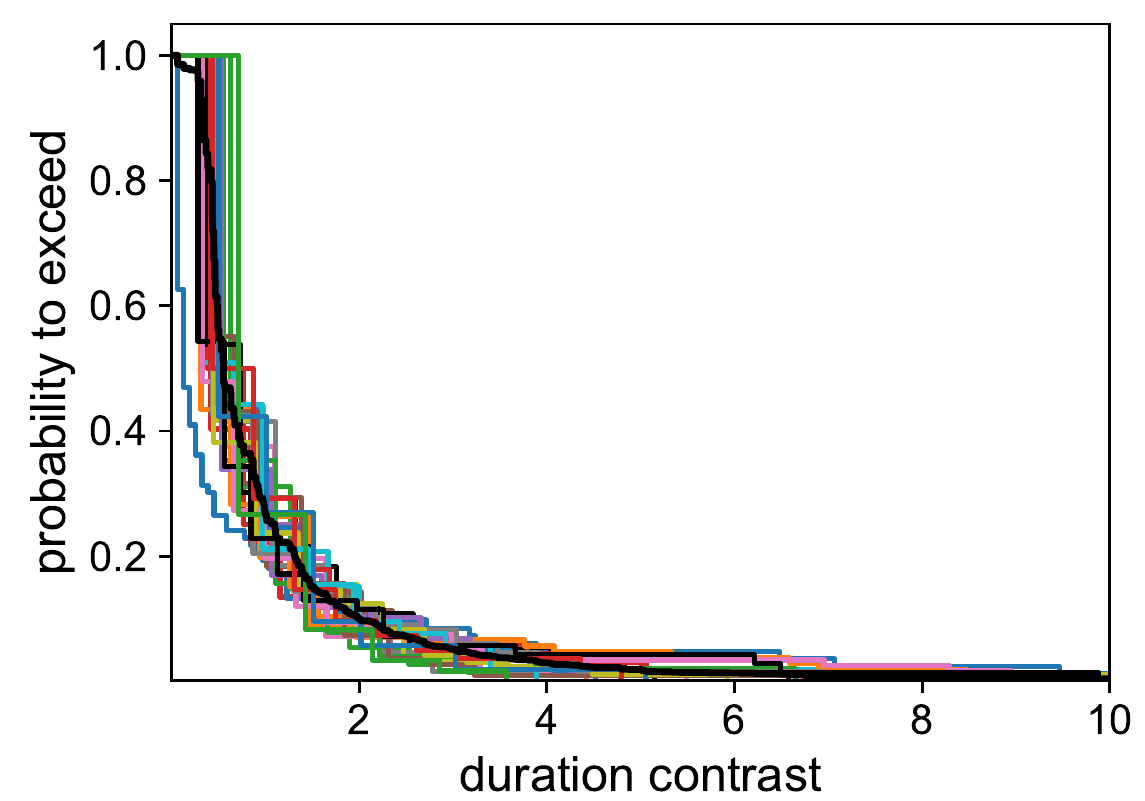}}
  \subfigure[\malawi]{\includegraphics[width=0.49\linewidth]{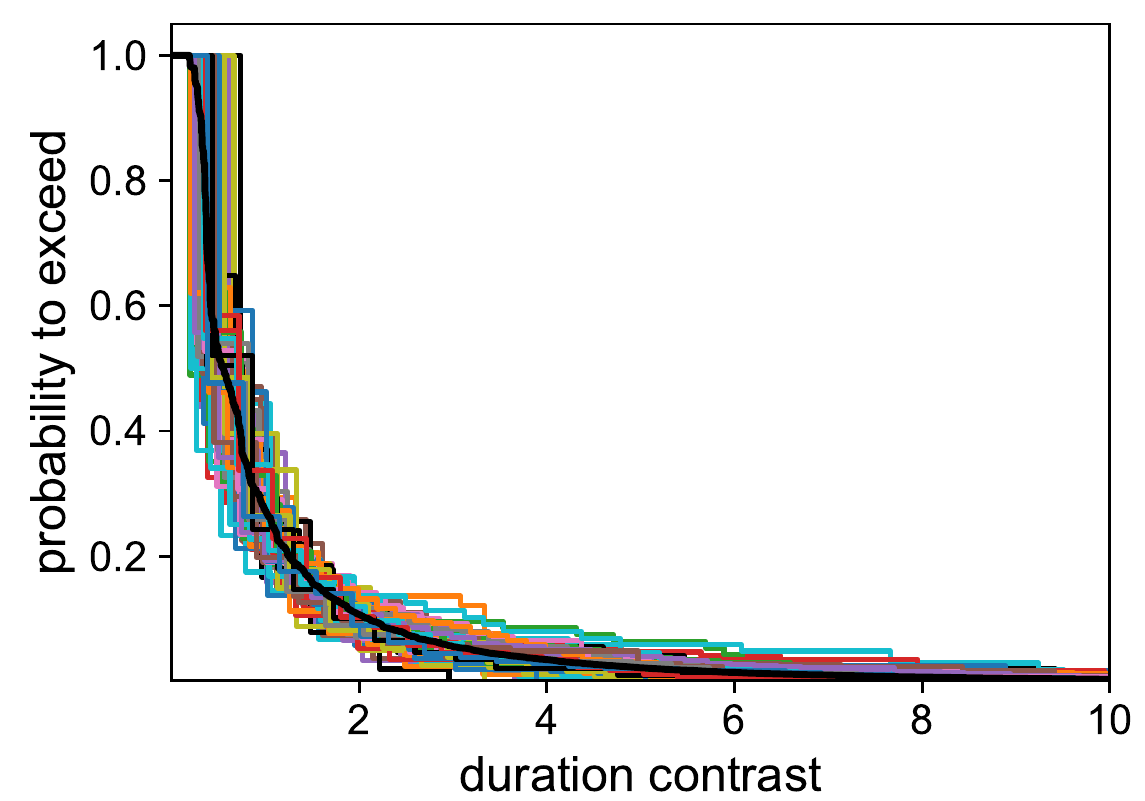}}
  \subfigure[\baboons]{\includegraphics[width=0.49\linewidth]{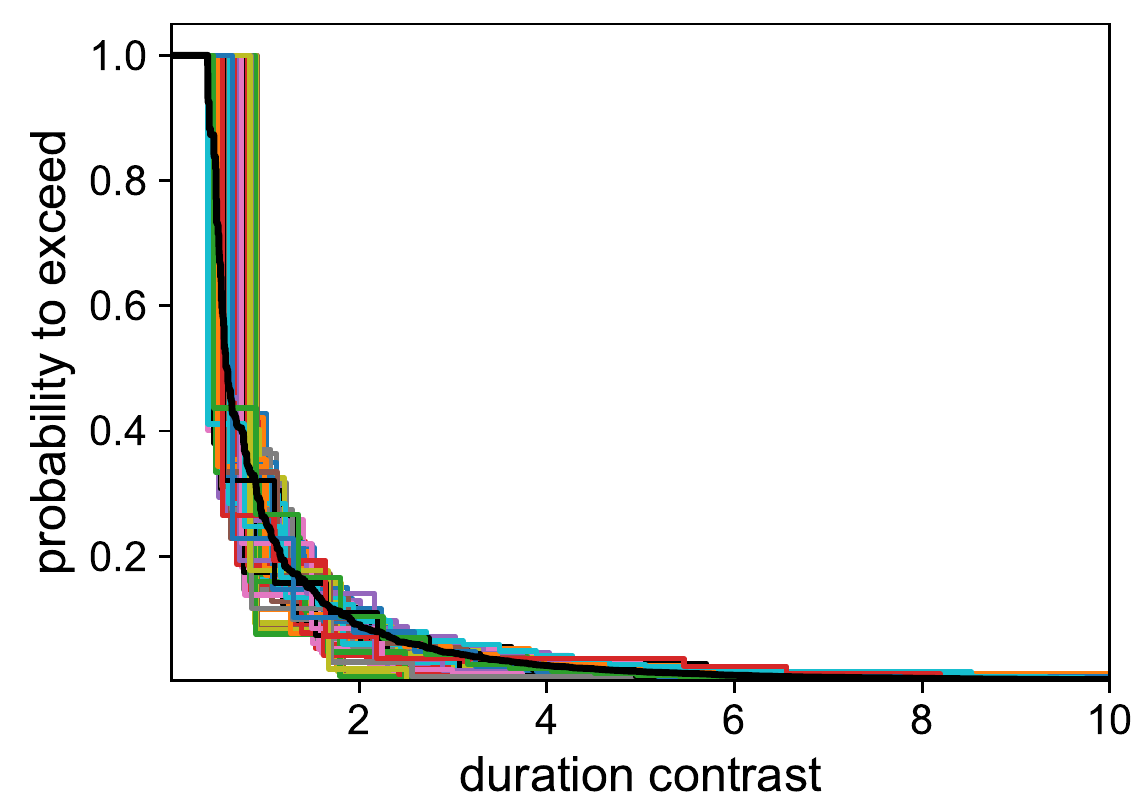}}
\caption{\label{fig:each_cdc} Distributions (\pte) of the contact duration contrast for
  each relation with at least 50 contacts. Each color represent a
  different distribution. The black line is the combined \pte shown 
  in \Fig{cdc_all}.}
\label{fig:cdc}
\end{figure}

At this point we have shown that the \textit{combined} contrast
duration (i.e. for all relations) follows a very similar distribution.
We now consider each relation separately and show in \Fig{each_cdc} a
superposition of the contrast duration distributions with the
$\Ncon(r)>50$ cut (similar results are observed without it but are
as expected more noisy, see \Sup{S4.2}).
They all follow rather closely the combined distribution, 
meaning that each relation is governed by the same universal rule.
Without the cut, our simulations still reproduces correctly the spread
observed for each relation (\Sup{S4.2}).

We thus arrive at the astonishing conclusion, that our \textit{individual} tendency to deviate
from the mean time spent in a face-to-face relation with a
given agent is essentially universal; it is the same in an occidental
hospital ward or conference, and in a village in Malawi. It is
also the same for social interactions among baboons in an enclosure. 

\section{Relation as a Levy process}
\label{sec:model}

We wish to model a relation as a random process, i.e the interaction between
\textit{a pair} of individuals by a \textit{single} mechanism. A relation is
an abstract concept and we do not give in the following a 
rigorous derivation of our model but rather highlight some of the
intuitions that lead to it.
Generally speaking a relation is sometimes ``tight'' and sometimes ``loose''. We
think about a correlated process where points would gather
for some time (the contacts) and then dissociate until the next gathering
(encounter). But we are not speaking here about
individuals evolving in the physical space who meet sometimes (as in \cite{Starnini:2013,Flores:2018})
but about a single point process evolving in an abstract space of ``relation''.
Since a relation involves two individuals, we will consider a 
space of dimension 2, challenging this hypothesis in the Discussion. 
The contrast is a dimensionless quantity; it does not depend on the
unit being used for the measurement of the contact duration. 
If the sensor resolution was different,
say T=1 or 10 s, the contrast distribution would still look the same
since we divide by the mean value (up to finite size effects). This
is a signature of a scale-invariant process.
We are then led to think about a Levy process or more precisely a Levy
flight (or ``walk'').

A Levy flight \cite{Mandelbrot:1975,Mandelbrot:1983} is a continuous random walk process where each increment
is isotropic and follows a radial distribution with a \pte 
$P(>r)=\dfrac{1}{r^\alpha}$ for $r>1$, where $0<\alpha<2$ is the Levy index.
Its particularity is that each step has
an infinite variance and that the point-process auto-correlation function is
a power-law of the distance between the points (for a precise
derivation in a 2D space, see \cite[Appendix A]{Plaszczynski:2022}). 
The stochastic path consists generally in several close
points followed by a long jump, with some further possibility to come
back near some previous samples.

\begin{figure}[h!]
  \centering
  \includegraphics[width=\linewidth]{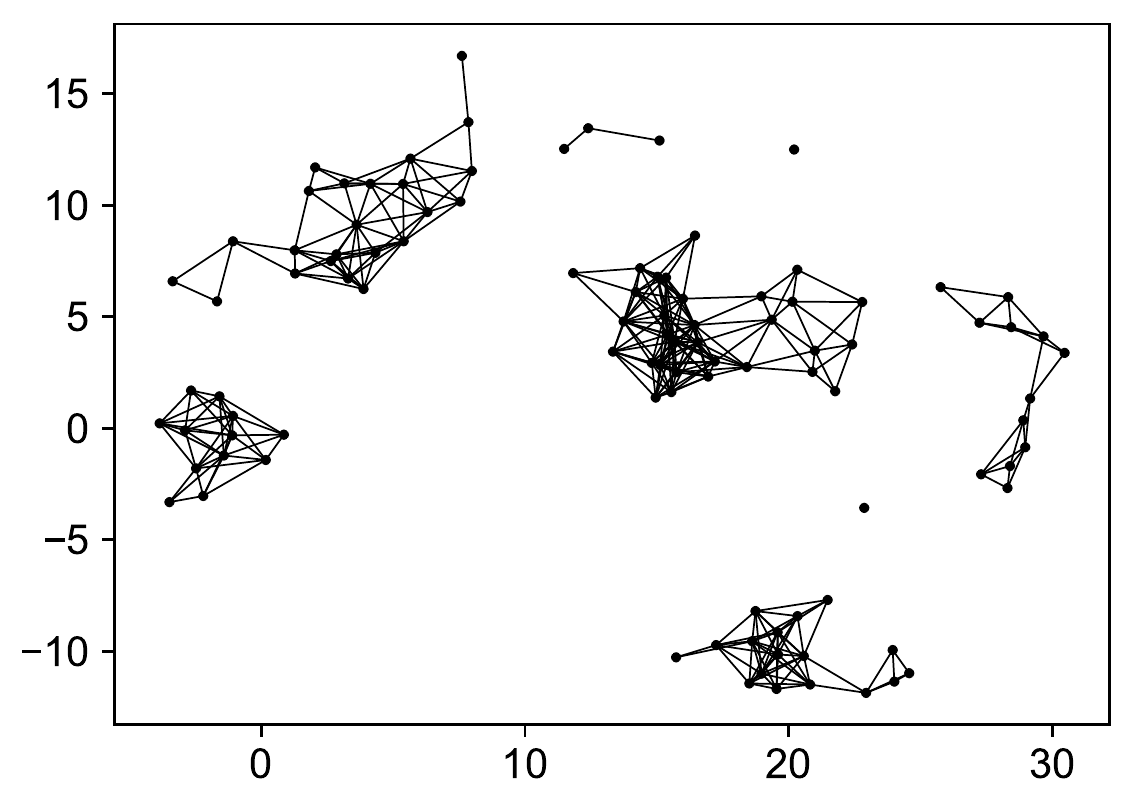}
  \caption{Example of a Levy Geometric Graph ($N=100$). Vertices 
    are placed at the position of a random Levy flight (here of index
    $\alpha=1$) and an edge is created if the
    distance between 2 vertices is below some distance (the
    scale, here 3).
    In our model each cluster represent an interaction between a given  pair of
    individuals, and the size, its duration (in units of the
    resolution steps).}
  \label{fig:LGG}
\end{figure}

If we connect all pairs of points that are below some given distance,
i.e. set some scale, we obtain a Levy Geometric Graph (LGG) whose properties were 
studied in \cite{Plaszczynski:2022}. An example is shown on
\Fig{LGG}.

The graph consists of a set of clusters (connected components) of
various sizes $C_i$.
It was noticed that the \textit{normalized cluster size}
\begin{align}
 c_i=\dfrac{\Nclus}{N} C_i,
\end{align}
where \Nclus is the total number of clusters and $N$ the total number
of vertices, is scale-invariant, i.e. depends only on $\alpha$.

In fact, we notice that since $N$ is the sum of the clusters size
\begin{align}
\label{eq:nclus}
  \dfrac{\Nclus}{N}=\left(\dfrac{1}{\Nclus}\displaystyle{\sum_{i=1}^{\Nclus}
  C_i}\right)^{-1}=\dfrac{1}{\bar C},
\end{align}
the normalized cluster size
\begin{align}
  c_i=\dfrac{C_i}{\bar C}
\end{align}
is precisely the cluster size \textit{contrast}.

Intriguingly, the distribution of this contrast variable (Fig.8 of
\cite{Plaszczynski:2022}) resembles closely that of \Fig{cdc_all}, 
which is at the origin of our model.

We then model each relation by a LGG. Each cluster represents an
encounter between the pair of individuals and
it size the duration of the contact (in units of the resolution step), 
i.e., $t_i=C_i$. 

To run the model, we need to define the size of the graph ($N$)
and its scale ($s$).

From the identification of the cluster size to the contact duration
\begin{align}
\label{eq:Nr}
  N(r)=\sum_{i=1}^{\Nclus} C_i=\sum_{i=1}^{\Ncon(r)} t_i(r)= w(r)
\end{align}
where $w$ are the weights discussed in \sect{struct}. 

Since we claimed that the cluster size contrast is independent of
the scale, we could use a priori any value for it.
However this property is only true asymptotically ($N\to\infty$) and for
large scales ($s\gtrsim 2$).
In our datasets this is often not the case, so we define it more
precisely in the following way.
For LGGs, it was shown \cite{Plaszczynski:2022} that the mean
number of clusters scales as
\begin{align}
  \dfrac{\Nclus}{N}=\dfrac{A}{s^{\alpha_c}}
\end{align}
where $A$ and $\alpha_c$ depends only on the Levy index. Their 
exact values are obtained using LGG simulations. For $\alpha=1.1$,
which will be important later,  we measure
$A=0.75, \alpha_c=1.4$ (other values may be found in \Sup{S5}).
From \refeq{nclus}, the corresponding scale for a given relation is
\begin{align}
\label{eq:sr}
  s(r)=\left[A \bar t(r)\right]^{1/\alpha_c}.
\end{align}

In order to account for the quantization of the time measurement
(20 s steps) the weights
$w$ and mean-times $\bar t$ are expressed as the number of resolution
steps and are thus integers.

\begin{figure}[ht!]
  \centering
  \includegraphics[width=\linewidth]{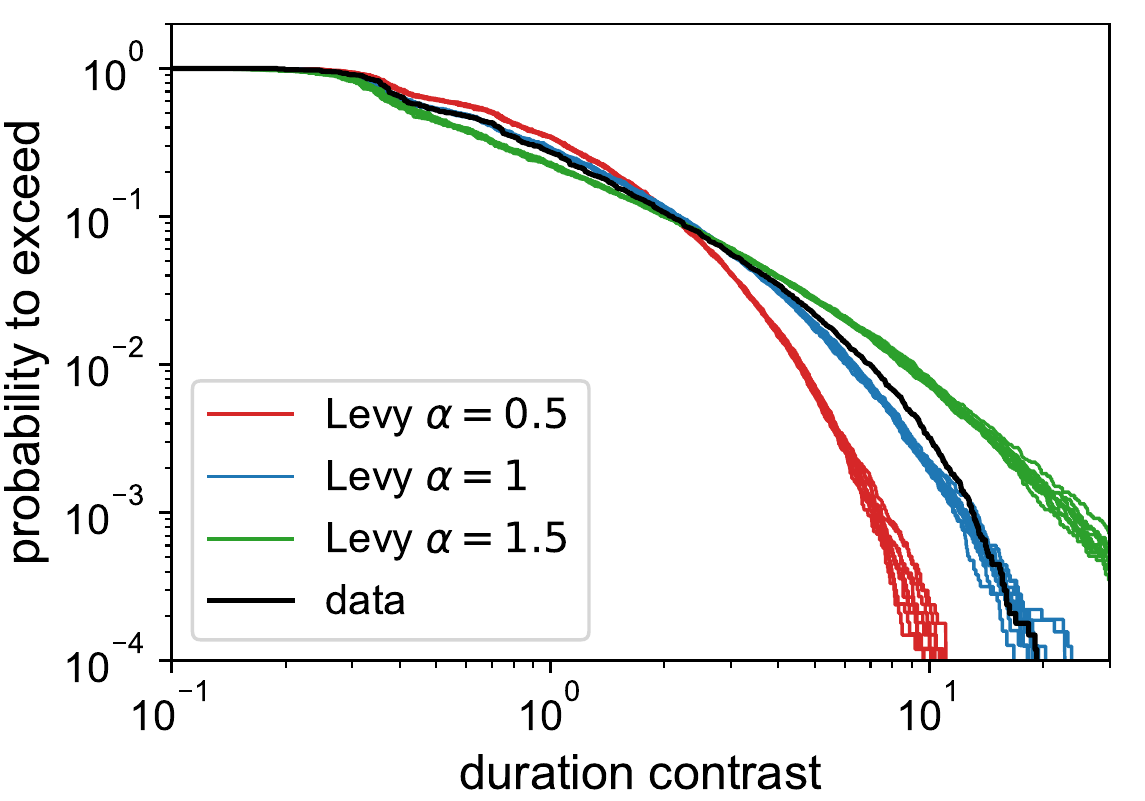}
\caption{\label{fig:Levy} Comparison between the 
  distribution of the duration contrast
  measured with the \malawi data (black line) and the cluster-size
  contrast obtained from 10 simulated LGGs per relation (colored
  lines) for 3 $\alpha$ values.}
\end{figure}

Each relation in a dataset is then modeled by an independent 
Levy graph of size given by \refeq{Nr} and scale given by \refeq{sr}.
Since our aim is to model the underlying contrast dynamics, we remove the
noise with the $\Ncon(r)>50$ cut.
LGGs being random graphs, we simulate 10 of them for each relation.
Each time we determine the connected components 
and compute their contrast size ($C_i/\bar C$) that we compare to the
contact duration contrast measured on the data. 
We illustrate how the result vary with some values of $\alpha$ in \Fig{Levy}.

\begin{figure}[ht!]
  \centering
  \subfigure[\hosp]{\includegraphics[width=0.49\linewidth]{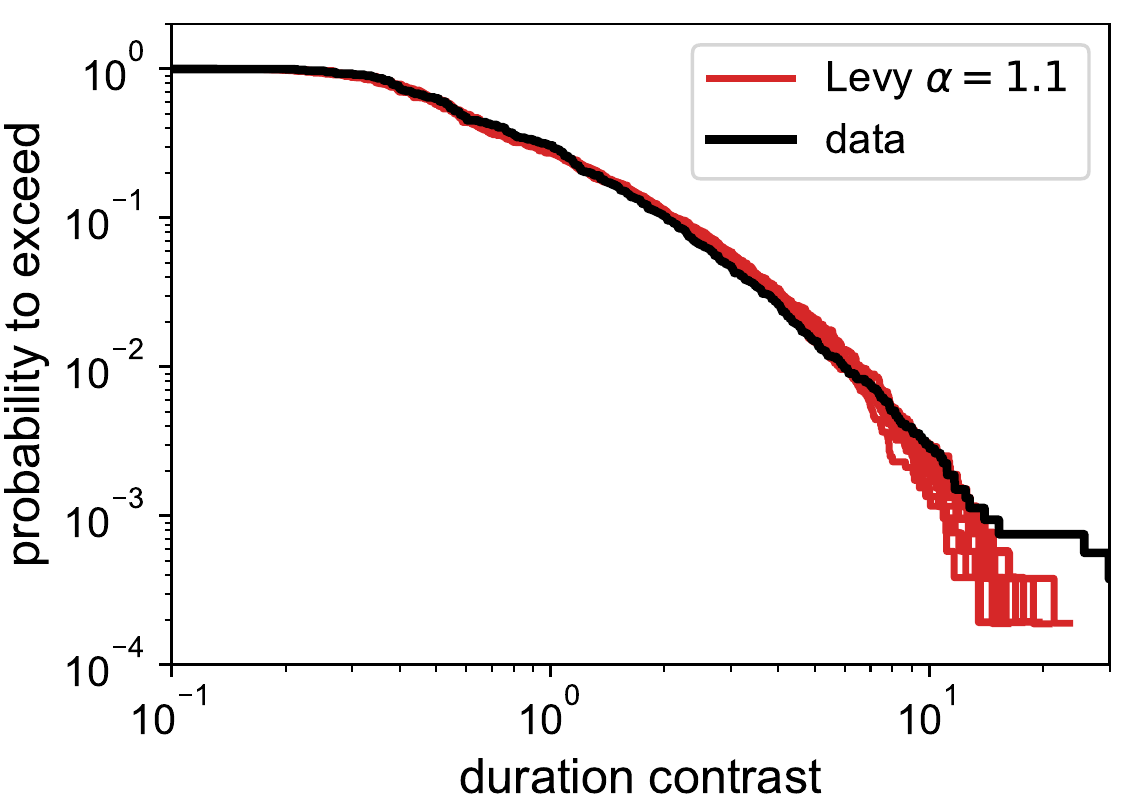}}
  \subfigure[\conf]{\includegraphics[width=0.49\linewidth]{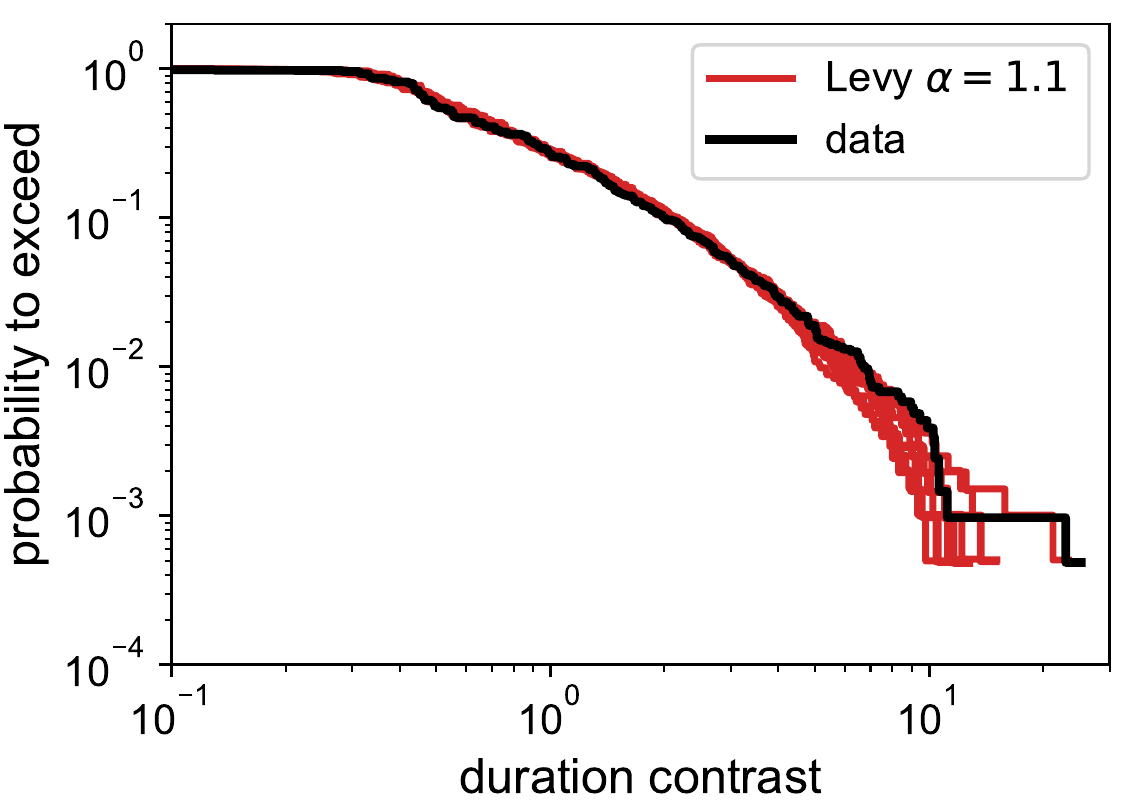}}\\
  \subfigure[\malawi]{\includegraphics[width=0.49\linewidth]{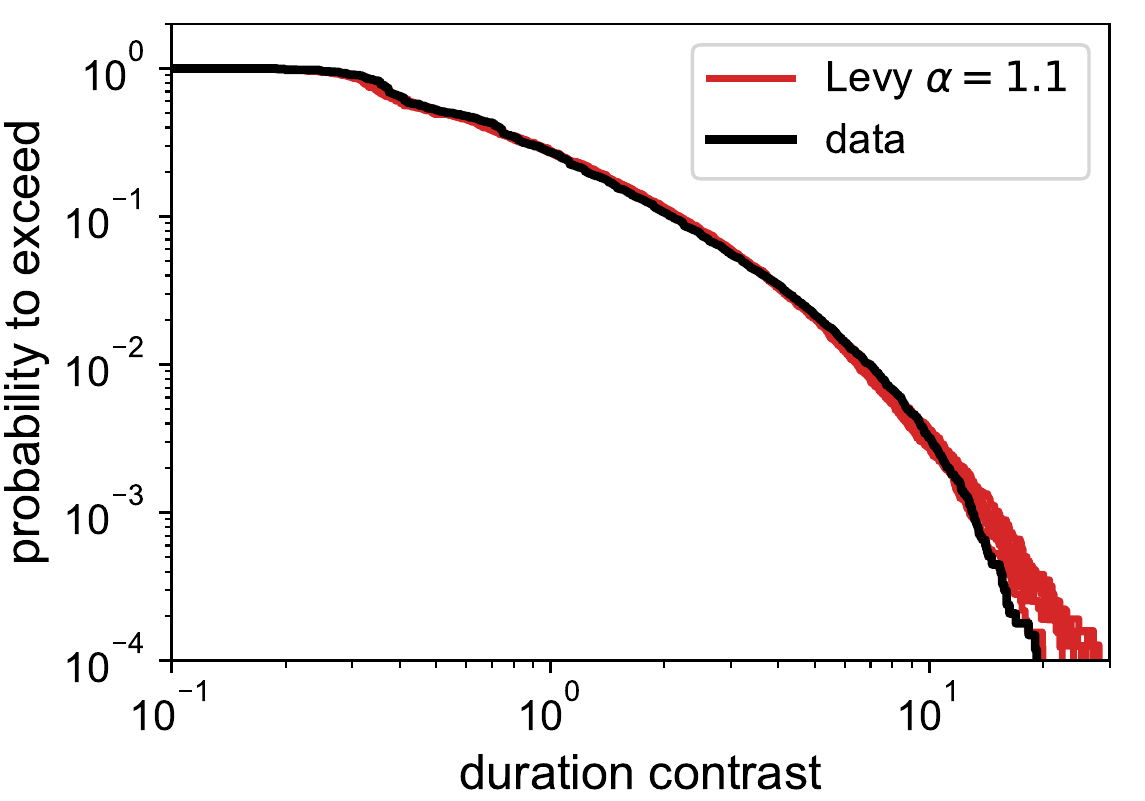}}
  \subfigure[\baboons]{\includegraphics[width=0.49\linewidth]{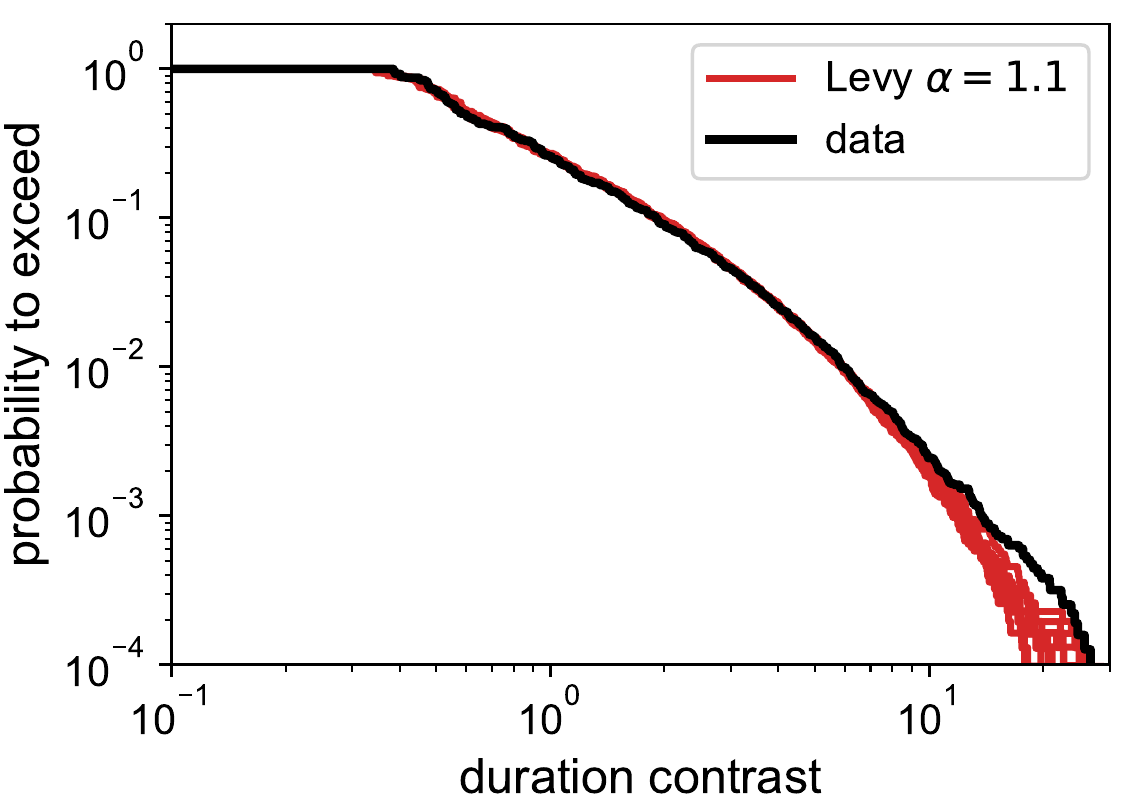}}
\caption{\label{fig:Levy_all} Comparison between the contrast data and
our model based on Levy Graphs with an index $\alpha=1.1$ on the 4
datasets. By comparing the \malawi result to \Fig{Levy}, we see
that an index of $\alpha=1.1$ reproduces slightly better the data than $\alpha=1$.}
\end{figure}

The best results are obtained with $\alpha=1.1$ and are shown in 
\Fig{Levy_all}.
Although data distributions have similar shapes (they correspond to
\Fig{cdc_all}(a)) the size $N(r)$ and scale $s(r)$ of the graphs 
are different for each dataset.
The agreement in all the cases is excellent.
There are small fluctuations among the random graphs and even the low
contrast wiggles are well reproduced.

\section{Discussion}
\label{sec:discussion}

We have compared face-to-face interaction data taken in some very 
different environments; some were recorded in an european hospital
or during a scientific conference, others in a small village in Africa.
We have also included data on baboons' interactions originally intended for
ethological studies, with the initial thought that they might serve as
null-tests for human behavior.
We recall that what we call ``interaction'' here is a face-to-face contact
within $\lesssim 1.5$ m and that lasts at least 20 s.
The kind of social exchange during this period is of course very different between
humans and animals. For the former it is most probably speaking, while
for baboons mostly groom and rest together \cite{Gelardi:2020}. 

Concerning the contacts between individuals, people in the \malawi
dataset interact with a smaller number of persons than in the occidental datasets
(hospital and conference). However the interactions are longer, 
so that the strength of the relations is similar. Baboons on the
contrary interact all with each other but for a shorter amount of time. This is
limited by the small number of animals (13) and it is not clear how
it would extrapolate to a number similar to the human experiments 
(around 80). This shows the high variability of social face-to-face
interactions between humans and also between animals.

Focusing on the temporal aspects of each interaction, it appears that the rate
and the mean-time spent together also depends on the social structure
(\Fig{tij}). 
However, once a contact is established, our finding is that the
deviation of the contact duration from its mean-value (the contrast) is almost
independent of the social structure (\Fig{cdc_all}); the over-duration
seems universal,
at least on these extremely heterogeneous datasets. 
But our main finding is that this general behavior is not
only valid on the whole population but also at the level of each individual
pair.
Since the interactions between the individuals are very different
(\sect{struct}), this indicates that the
amount of extra-time spent interacting with a given individual is not
strongly affected by the social environment.
Once a relation is triggered it has its own dynamics. It can be
factored-out from the sociological complicated structure.
While one may think that our ability to interact more or less 
with a given person is related to a common human level of ``patience'', 
the same results for baboons shed doubt on this explanation.

The shape of the contrast distribution mostly comes from the one of the
interaction duration (\Fig{tij}). Normalizing it by each relation
mean-time (\Fig{Ncon}(b)) \textit{standardizes} it to a common distribution. 
This is not a trivial result obtained by dividing by the mean value. 
When there are many samples ($\Ncon>50$), the arithmetic mean converges to a
single number (the statistical mean) and becomes mostly independent of
the individual samples. The contrast distribution then follows
essentially a rescaled version of the duration one. If the contact duration
were to follow for instance a Poisson distribution, the contrast would have been very
different, going to 0 around $\delta=4$ (see \Sup{S6}).
We recall that the scaling varies with the relation since it depends on each mean-time value.

This ``universality'' that we find on four very different datasets, 
holds for face-to-face interactions ($\lesssim1.5$ m for at least 20 s) 
between adults in a mildly dense closed
environment. This is observed up to a contrast of about 10. Some
differences seem to emerge for larger values 
but it is difficult to be certain given the data we have at
hand. We emphasize that our main finding is that \textit{each} relation follows
a similar pattern.
This must be put to test with other data. Although they probe a
different kind of social interactions, mobile phone communications
could show some similar characteristic. 

Since there is a clearly observable curvature in \Fig{cdc_all}(a) the contrast distribution
does \textit{not} follow  a power-law. Neither is it of exponential type. 
We obtain a good fit to the \pte in the $0.4<\delta<10$ 
region with the following function
\begin{align}
  \label{eq:fit}
  P(>\delta)=0.37 e^{-0.3\delta}/\delta.
\end{align}

\begin{figure}[h!]
  \centering
  \includegraphics[width=\linewidth]{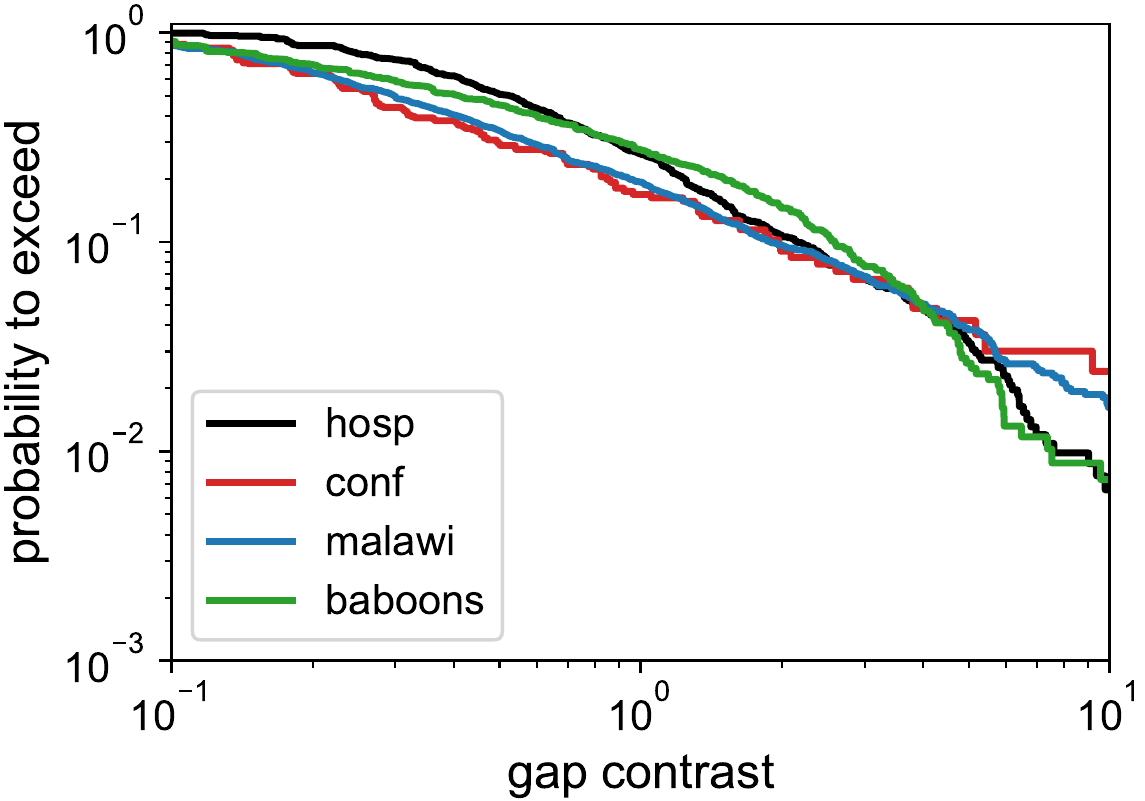}
  \caption{Distributions of the contrasts of the gap-time
    (inter-contact duration) on our datasets. 
    To avoid the long night breaks, we show results for a single day.}
  \label{fig:gap}
\end{figure}

We have considered so far the duration of the interactions but not 
\textit{when} they happen, i.e. the inter-contact or ``gap'' time.
Their contrast could also show some universality. This is not the case as
shown in \Fig{gap}. The contrast of the inter-contact time depends on
the sociological environment.

Some models that describe face-to-face interactions do exist.
They are either based on a mechanism of preferential attachment to
groups of individuals
\cite{Stehle:2010,Zhao:2011} or on a (standard) random walk that 
describes the physical displacements that must be biased
\cite{Starnini:2013,Sekara:2016,Flores:2018}. 
They all contain parameters to be adjusted, while our model has none
once $\alpha$ is fixed to 1.1. 
Whether the agent-based models can reproduce the contrast for \textit{each} relation, in particular
on the \malawi and \baboons datasets, remains to be proven.
But the essential difference with our model is that they are all 
\textit{agent-based}, trying to reproduce a collective behavior from
each individual one, while our approach is
to model the very concept of \textit{relation}.
In some aspects, agent-based models have a richer phenomenology 
since they can predict the structure of interactions between individuals
while we use as an input the distributions of the interaction rates
$\Nint(r)$ and the mean-time $\bar t(r)$.
Agent-based models could then be used to define those values, handling the detail
of the relation within our model, i.e separating the ``sociological''
part from the ``universal'' one. 

Power-law distributions are ubiquitous in the field of human interactions.
They are observed for instance in the delay to answer e-mails
\cite{Barabasi:2005}, in human travels \cite{Brockmann:2006}, or 
on the time we spend at a given location \cite{Song:2010}.
Interestingly, the measured power slopes are always around 1.
But they are obtained very differently, by fitting 1D temporal data
assuming a power-law distribution. 
In our model the Levy index $\alpha$ characterizes a random-walk in 
2D space, and does not lead to power-law distributions.

The idea behind using Levy graphs is the following.
In a particular relation, the duration of the contact has no particular
scale; it essentially depends on our free-will.
However due to social constraints, especially for work-related
activities, there exist some implicit limitation on the length of each
discussion so that the mean-time in each relation does not drift too
much (no one spends the full day talking with the same person).
This sets a scale, that can be different for each pair according to
the level both participants enjoy interacting.
Levy graphs precisely describe what happens when we apply a scale
(here related to $\bar t$, see \refeq{sr}) to a scale-free process (the time during which we communicate).
What appears immediately when studying LGGs, is that their clusters have a very
specific size distribution. In particular their
contrast is similar to the one obtained for the contact durations.
A closer examination revealed us that this agreement was
in fact excellent and that all the data can be reproduced very
accurately with a unique Levy index, $\alpha=1.1$.
The main argument in favor of this model is then its success for 
matching non-trivial data with a single parameter.

One may ask what the nature of the ``space of relation'' is.
We note that in social networks, the ``space of relations'' is
also an abstract concept. It is only used for representation, the real
mathematical object being the graph itself.
A difference is that a Levy flight requires a \textit{metric} space to
perform the steps. However the distance is dimensionless.
Indeed a genuine Levy flight has a minimal step size $r_0$. When connecting
points, the linking length $L$ has the same dimension. What we call
the scale is $s=L/r_0$ . In our model, without loss of
generality, we have assumed $r_0=1$. In
\Fig{LGG} we could have scaled as well the axes by a factor 2 using
$r_0=2$. As in social networks, what
really matters is the graph itself not its representation.

The 2D space is dictated by the data.
We have checked that using a Levy walk in
a higher dimension space, the agreement with the data gets worse (\Sup{S7}).
This is maybe due to the fact that a relation involves two agents. 
However this then raises another question.
In a low dimensional space, one cannot consider the Levy process as
solely consisting of a few local steps followed by a long jump leading
to a new cluster. There is a non-negligible probability (about 20\%
for $\alpha=1$, see \cite{Plaszczynski:2022}) that
the walker comes back to a previous set of points and contributes to the
cluster size, thus to the duration of the contact in our model. 
Increasing the dimension, the return probability decreases and
the contrast \pte converges to $e^{-\delta}$, which is not observed on
data (\refeq{fit}). The meaning of this return-probability, which is not
a physical displacement in the real space, remains to be understood.

Levy flights belong to a very peculiar class of correlated point-processes
without any mean-density but a power-law autocorrelation function of the
form $\xi(r)\propto 1/r^{2-\alpha}$ (in 2D) where $r>1$ is the dimensionless
distance in the relation space. Such an autocorrelation gives rise to
scale-invariance since for any $b$ we have $\xi(br)\propto \xi(r)$. The
process looks (statistically) the same at any scale.
This translates for Levy graphs into the fact that the cluster-size
contrast is scale-invariant. This is only true for a large number of
samples and for a scale above $\gtrsim 2$.
Both could be achieved with an experiment with a finer timing
resolution, since the time units ($t_i$ and $\bar t$) would increase.
A prediction from our model is that the contrast distribution
would not show anymore the wiggles at low contrast (\Sup{S8}). 

In classical random geometric graphs, based on uncorrelated points, 
a geometric phase transition appears when increasing the scale; most points
end up belonging to a single cluster, the giant component. In 2D this happens
around a mean degree of 4.5 \cite{Dall:2002}. This is not the case for Levy
graphs that escape this capture due to the long jumps.
This property leads in our case to the fact that clusters, i.e. 
encounters between two individuals, 
always exist whatever the
scale is. It is a necessary condition for a random-graph based model,
since otherwise, by increasing the resolution of the instrument
the size and scale (see \refeq{Nr},\refeq{sr}) would also
increase, leading to a single very long interaction.

Finally, Levy flights have been much discussed in the context of 
movements of foraging animals \cite{Raposo:2011} and were also observed 
in some human communities practicing hunting
\cite{Raichlen:2014} or looking for resources \cite{Reynolds:2018}.
Whether they really represent the optimal random-search strategy has been hotly debated 
\cite{Viswanathan:1999,Viswanathan:2008,Raposo:2009,James:2011,Reynolds:2015},
but they are certainly advantageous in several aspects 
\cite{Raposo:2003,lomholt:2008,Humphries:2014}.
This suggests that we may also be searching for something in a
face-to-face relation.  

\section*{Conflict of interest disclosure}
The authors declare no conflict of interest.

\section*{Data availability}
All data are available form the \textsl{sociopatterns} web site  \safeurl{www.sociopatterns.org}

\section*{Software availability}
The (python3) scripts used to produce the results are available from 
\safeurl{https://gitlab.in2p3.fr/plaszczy/coll}
(the entry point is \textsf{gtAgg.py}).

\section*{Acknowledgements}
The authors thank Basile Grammaticos and Mathilde Badoual for
  fruitful discussions and corrections on the manuscript. The 
  graph-related computations and \Fig{graphs} were performed using
  \textsf{graph-tools} : 
\safeurl{https://graph-tool.skewed.de}

% Bibliography

%\bibliographystyle{apsrev4-2}
\bibliographystyle{unsrtnat}

\bibliography{references}

%\widetext
%\clearpage
%\renewcommand{\thetable}{S\arabic{table}}
%\setcounter{section}{0}
%\renewcommand{\thesection}{S-\Roman{section}}

\end{document}

%%%%%%%%%%%%%%%%%%%%%%%%%%%%% Begin Document Here %%%%%%%%%%%%%%%%%%%%%
\bigskip
\title{ 
``How much do we stand our fellows? \\A universal behavior in
face-to-face relations''
\\\textbf{Supplementary Information}}
\author{ S. Plaszczynski and G. Nakamura}
\maketitle
\tableofcontents
\newpage

\section{Main results}
\label{sec:Main}

There are 2 major results in this work.

The first one concerns the duration of face-to-face contacts between
\textit{each} pair of individuals. Although
the mean value may vary, the \textit{deviation} from this mean value is very similar in several very different social 
contexts. As in other fields of physics (e..g cosmology) we call this
over/under-duration the duration \textit{contrast}. It is a dimensionless value
that describes how much given an interaction is longer or shorter than its
usual (mean) time and can be expressed  in percents. By construction its mean value is 1.

To remove some statistical noise , we only use data where there is a sufficient number
of interactions (samples) \Ncon. Since the duration distribution is heavy-tailed we
use a large minimal value of 50.

\begin{figure}[ht!]
  \centering
  \subfigure[]{\includegraphics[width=0.49\textwidth]{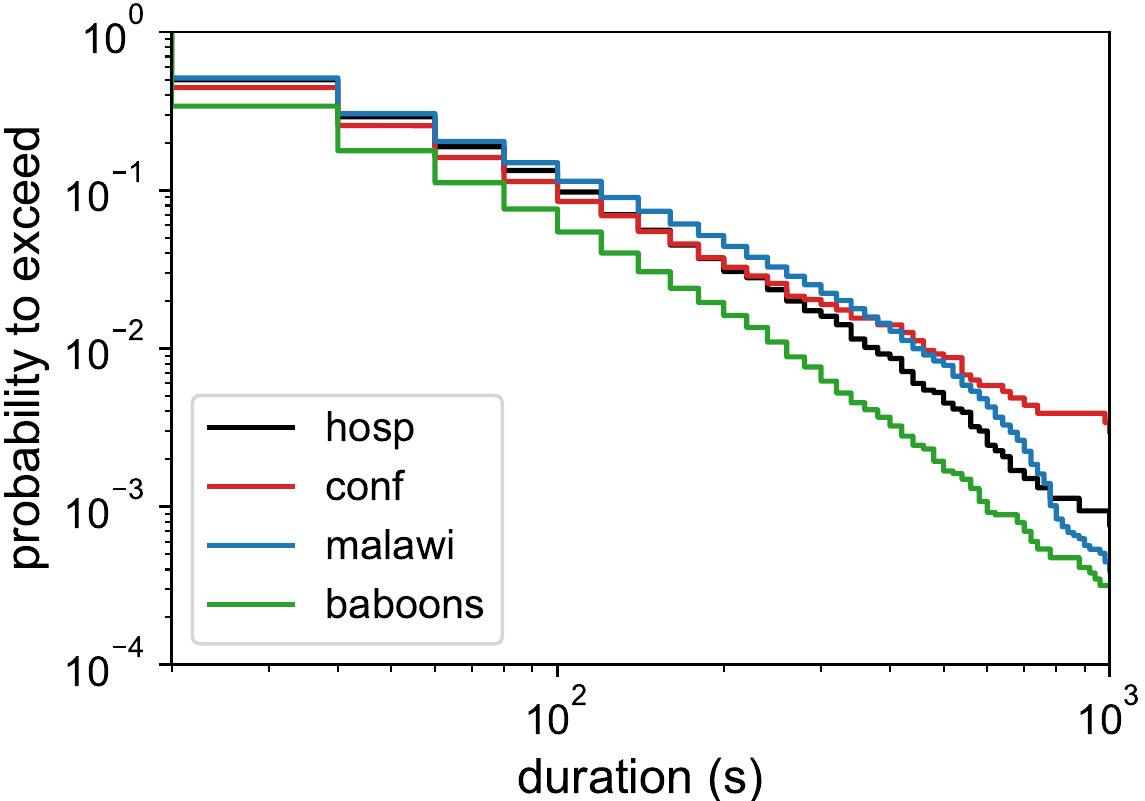}}
  \subfigure[]{\includegraphics[width=0.49\textwidth]{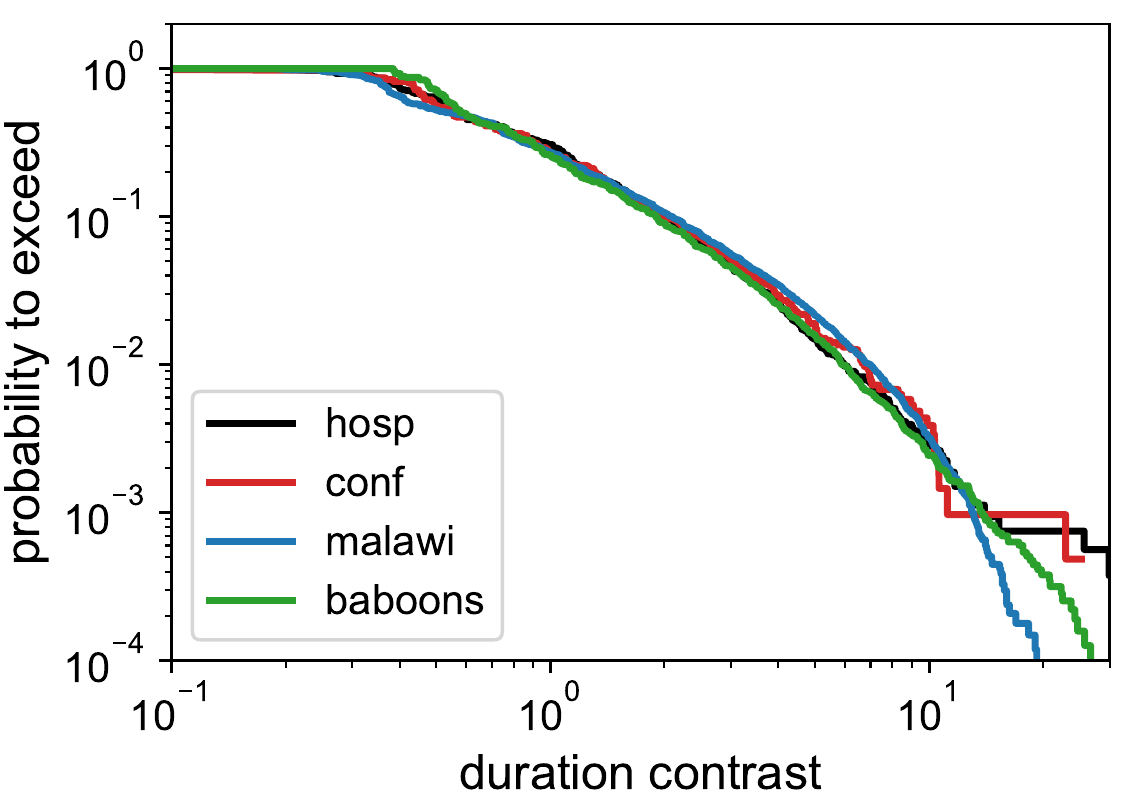}}
\caption{\label{fig:cdc} (a) distributions (\pte) of the durations of contact and (b) contrast
durations on the 4 datasets. The cut $\Ncon>50$ is used for both.}
\end{figure}

\Fig{cdc} (a) shows how the distribution of the duration of contacts
and how it changes when computing the contrast in (b). Both use the
$\Ncon>50$ cut.

The second major result is the  spectacular agreement of our model with
the data. We recall how the model proceeds:
\begin{enumerate}
\item on a given dataset we loop on all the pair of contacts that were
  registered (``relation'', $r$) over the full duration of the
  experiment. We keep only timelines where there are at least 50
  intervals of interaction.
\item for a given pair ($r$) we determine its mean duration value :
  $\bar t(r)$.
\item we draw 10 random Levy graph with $\alpha=1.1$ for each 
  relation $r$. Their size is given by the weight $N(r)=w(r)$,
  and their scale (linking
  length) by $s(r)=(A \bar t(r))^{1/\alpha_c}$, where  $A\simeq 1$ and
  $\alpha_c\simeq \alpha$ are
  determined from LGG simulations and are given for some $\alpha$
  values in \sect{A}. Times ($w$ and $\bar t$) are expressed as a raw
  number of resolution steps (20 s) and are thus integers.
\end{enumerate}

Although the contrast data are very similar (\Fig{cdc} (b)) we show
the results on the 4 datasets since the sizes $N(r)$ and scales $s(r)$
are different in each case.

\begin{figure}[ht!]
  \centering
  \subfigure[\hosp]{\includegraphics[width=0.49\textwidth]{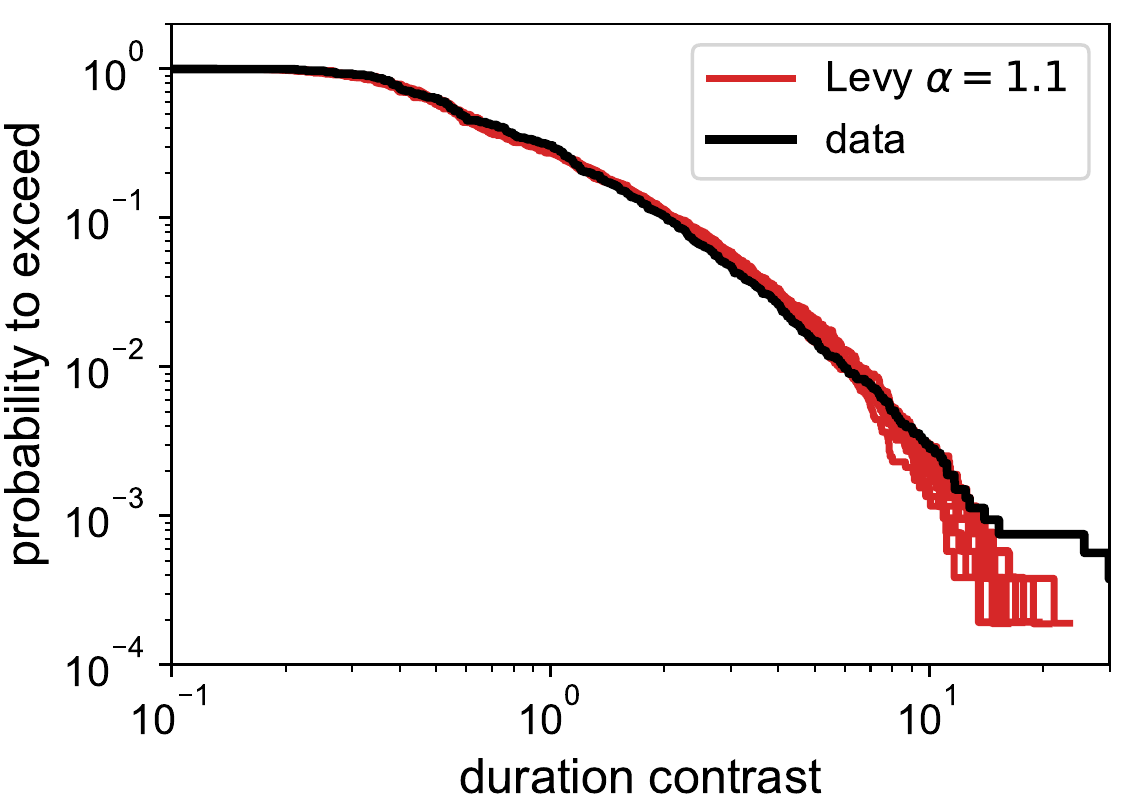}}
  \subfigure[\conf]{\includegraphics[width=0.49\textwidth]{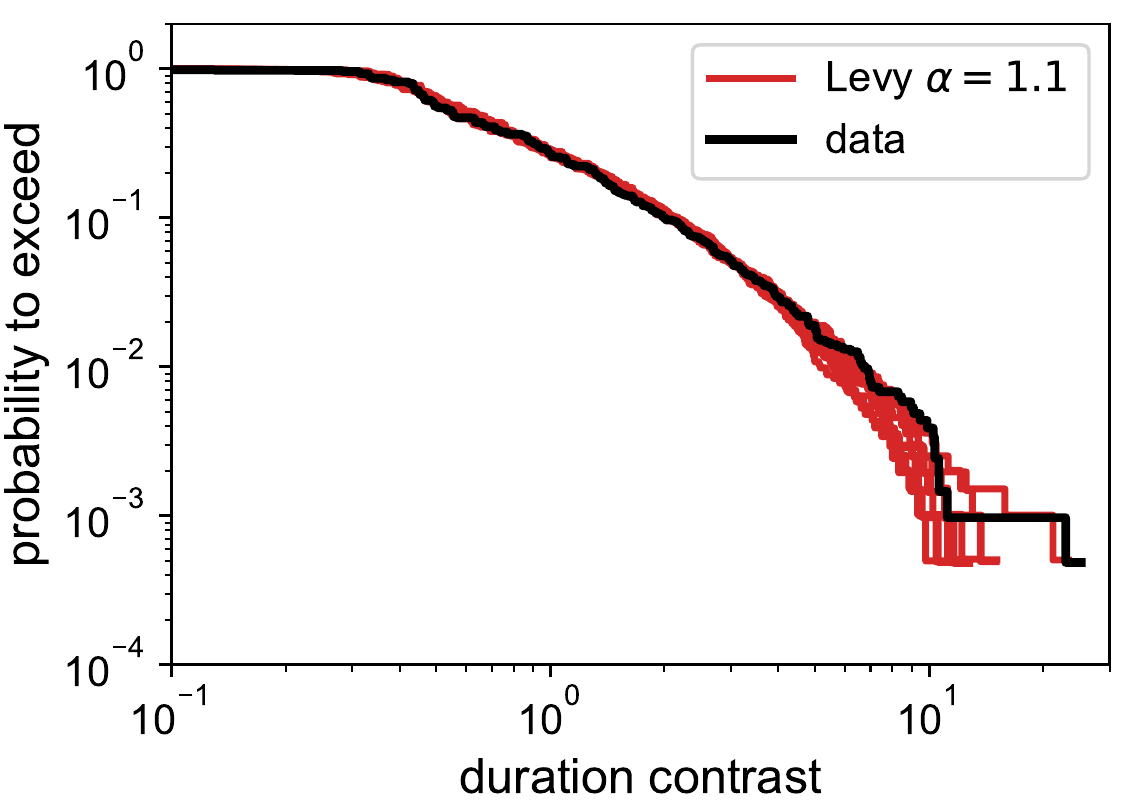}}
  \subfigure[\malawi]{\includegraphics[width=0.49\textwidth]{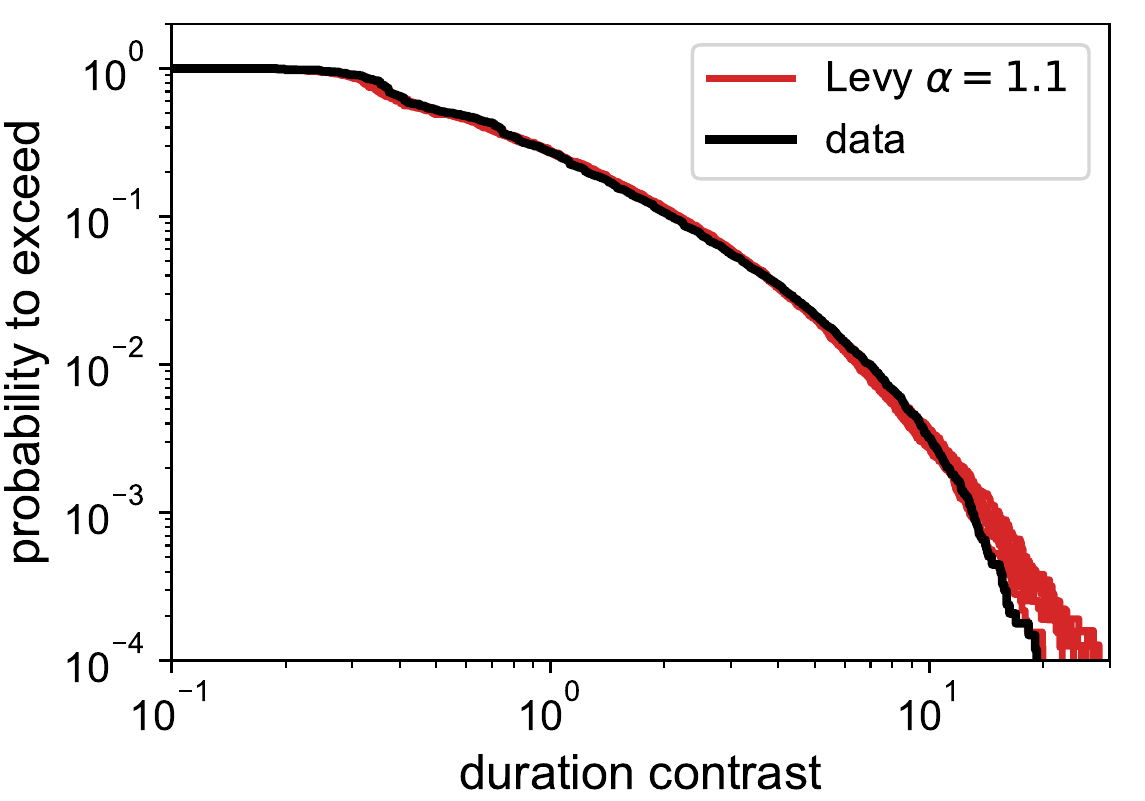}}
  \subfigure[\baboons]{\includegraphics[width=0.49\textwidth]{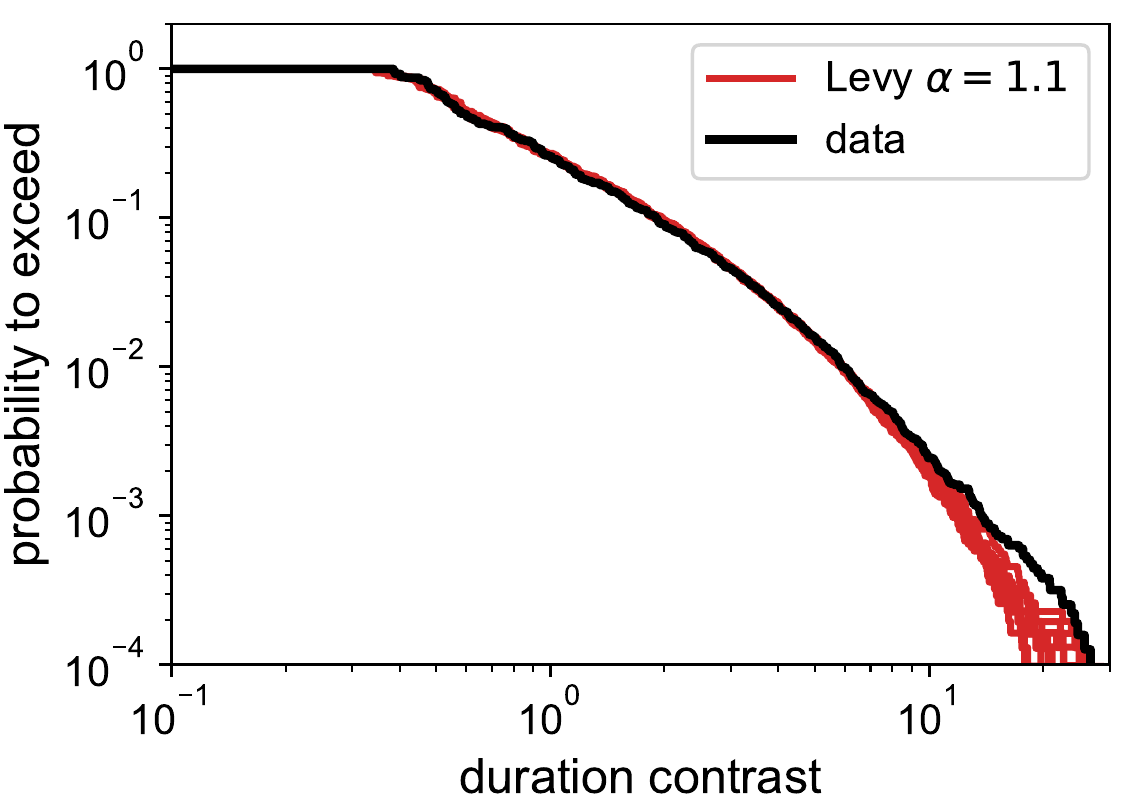}}
\caption{\label{fig:models} (a) distributions (\pte) of the durations of contact and (b) contrast
durations on the 4 datasets. The cut $\Ncon>50$ is used for both.}
\end{figure}

\section{Other datasets}
Although in our work we have focused on the sociologically 
different datasets, we have also looked at some other data
provided by the \textit{sociopattern} collaboration.
\begin{enumerate}
\item \sfhh: these are data taken at another conference (SFHH,\cite{Cattuto:2010,Stehle:2011})
with \about 380 participants for 2 days.
\item \office:  data taken at an office (Intitut de Veille Sanitaire
  near Paris,
  \cite{Genois:2015,Genois:2018}) with about 165 participants for 10
  days.
\item \highschool: data from a french high-school, near Marseille  \cite{Fournet:2014,Mastrandrea:2015}.
  About 300 participants for 4 days 
\end{enumerate}

\begin{figure}[ht!]
  \centering
  \subfigure[]{\includegraphics[width=0.49\textwidth]{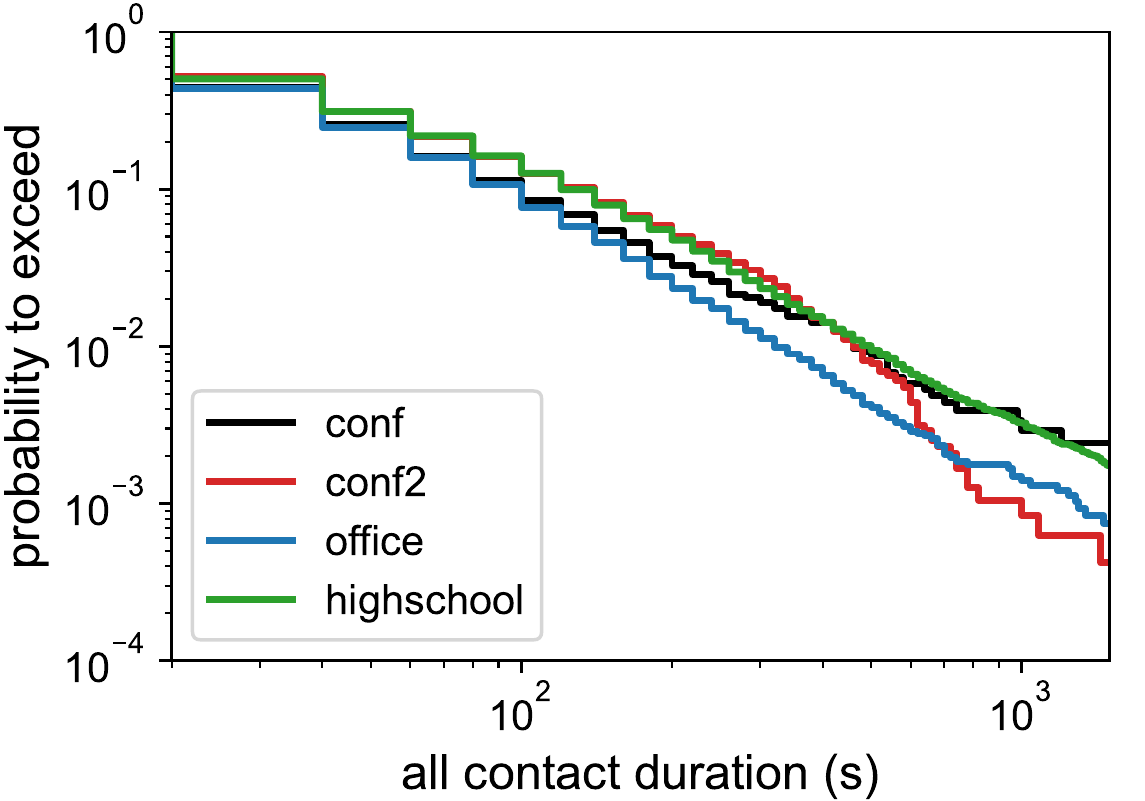}}
  \subfigure[]{\includegraphics[width=0.49\textwidth]{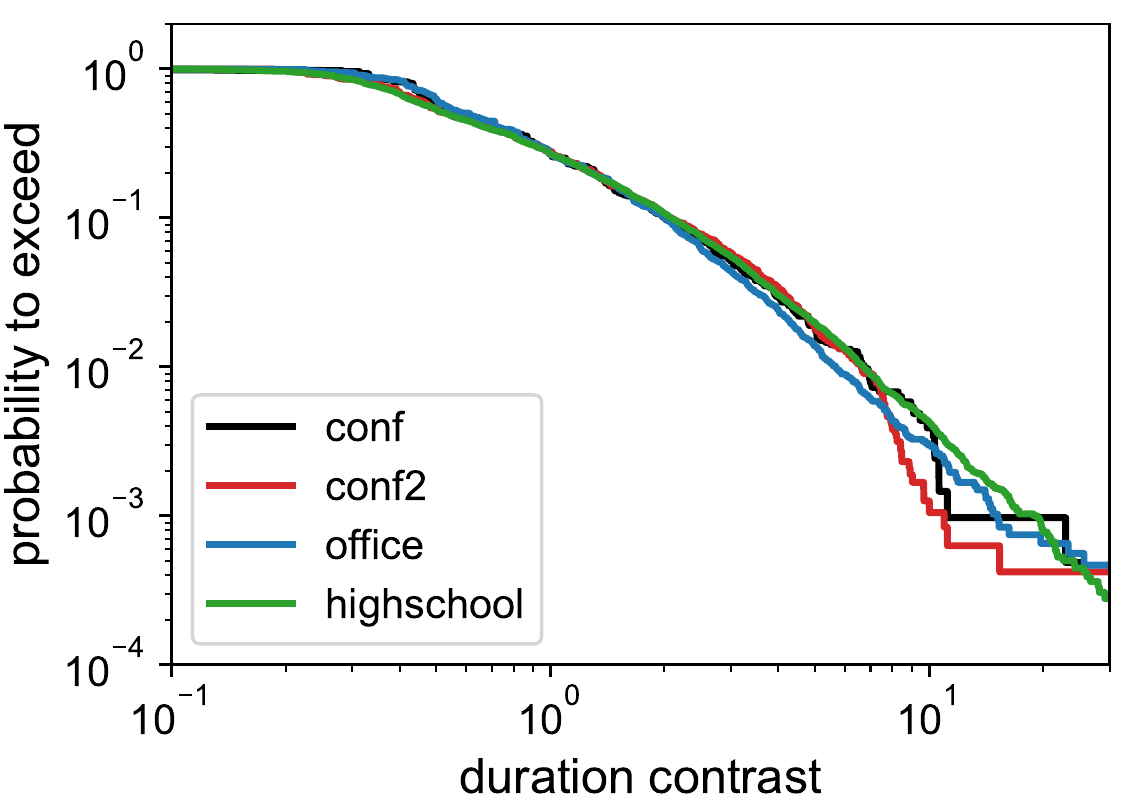}}
\caption{\label{fig:cdc_new} (a) distributions (\pte) of the durations of contact and (b) contrast
durations for 3 other datasets.}
\end{figure}

As in \sect{Main} we show how using the contrast standardizes the
contact durations on \Fig{cdc_new} on these new datasets; for comparison we
also included the previous \conf result and still use the $\Ncon>50$ cut.

The contrast distributions are similar to the results presented in
the paper, and differences are barely noticeable in a linear representation.

\begin{figure}[ht!]
  \centering
  \subfigure[]{\includegraphics[width=.7\textwidth]{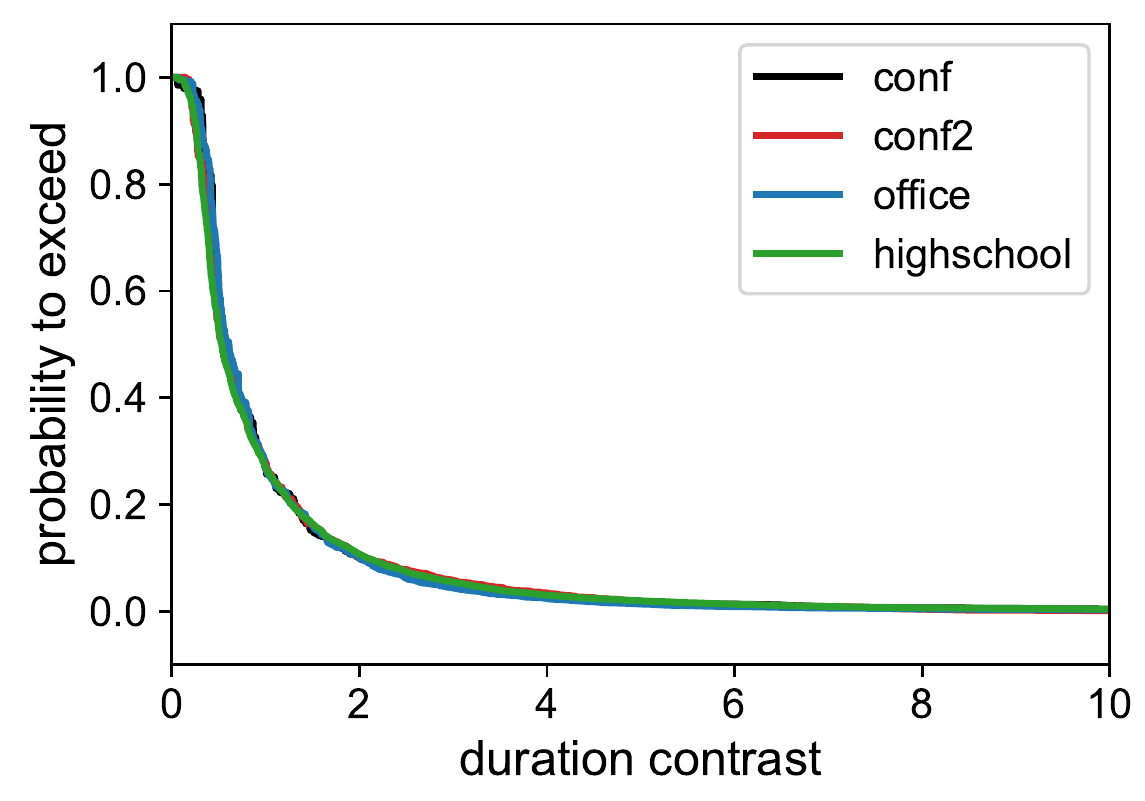}}
\caption{\label{fig:cdc_new_lin} Same as \Fig{cdc_new} (b) in linear scale.}
\end{figure}

\section{Number of interactions}

\begin{figure}[ht!]
  \centering
  \includegraphics[width=0.7\textwidth]{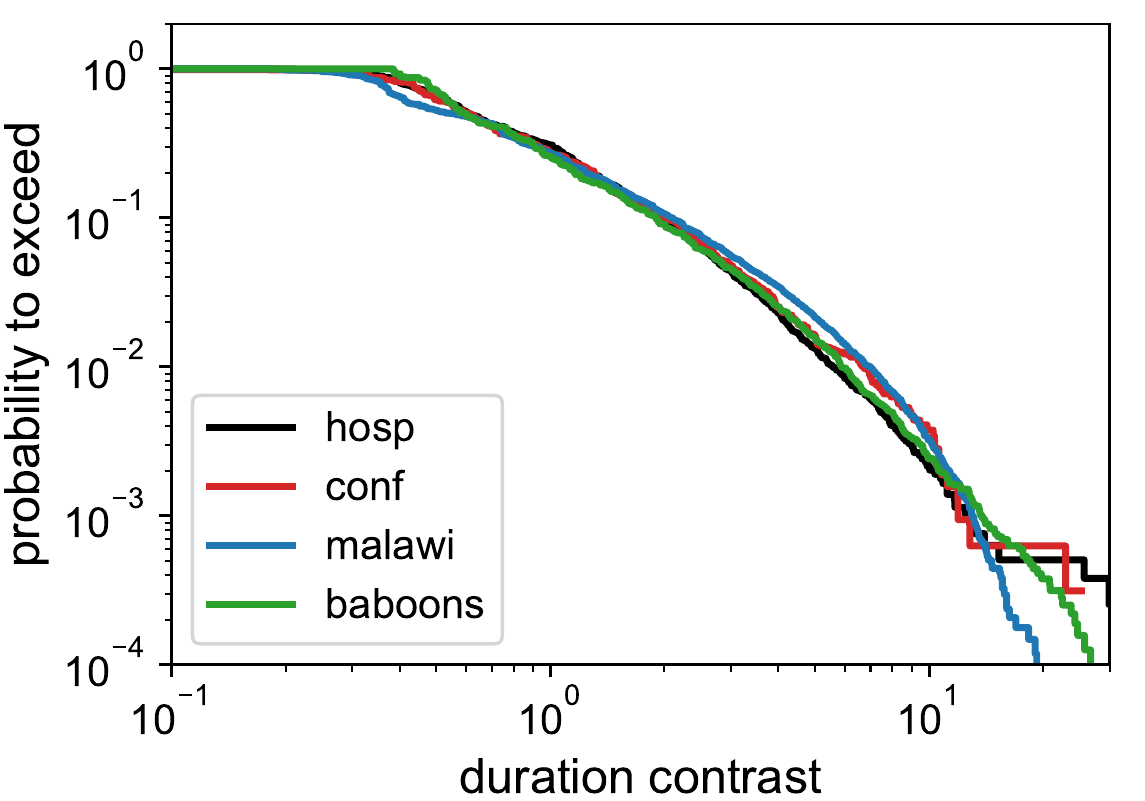}
\caption{\label{fig:Nint30} Distribution of the contrast durations on
  the 4 datasets  using a $\Nint(r)>30$ cut.}
\end{figure}

To compute a mean value one (traditionally) invokes the central limit
theorem and compute a reliable arithmetic mean with $\simeq 20$
samples . However the contact duration distribution is very wide
(\Fig{cdc}(a)) so we prefer to increase that value to $\Ncon>50$
before computing the contrast, still keeping a reasonable number of
timelines in each case.
We may use a lower cutoff and the contrasts distributions are still
similar(\Fig{Nint30}) . But we have added some noisy samples. Note that
this value of 50 is only dictated by the data limit data-taking period.
Had we a really long period, the mean interaction times between each
pair would be well known and this cut would be not necessary.

\section{Simulations without a minimal number of interactions}

\subsection{Combined contrast}

\begin{figure}[ht!]
  \centering
  \subfigure[\hosp]{\includegraphics[width=0.49\textwidth]{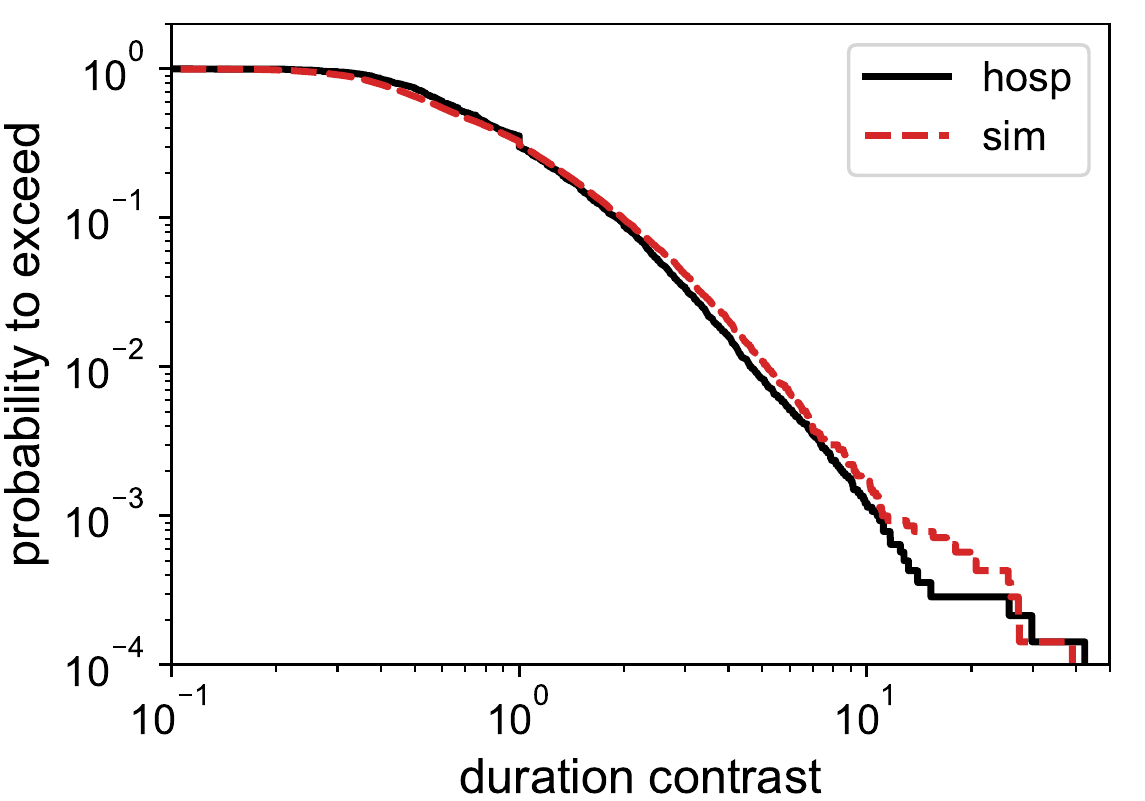}}
  \subfigure[\conf]{\includegraphics[width=0.49\textwidth]{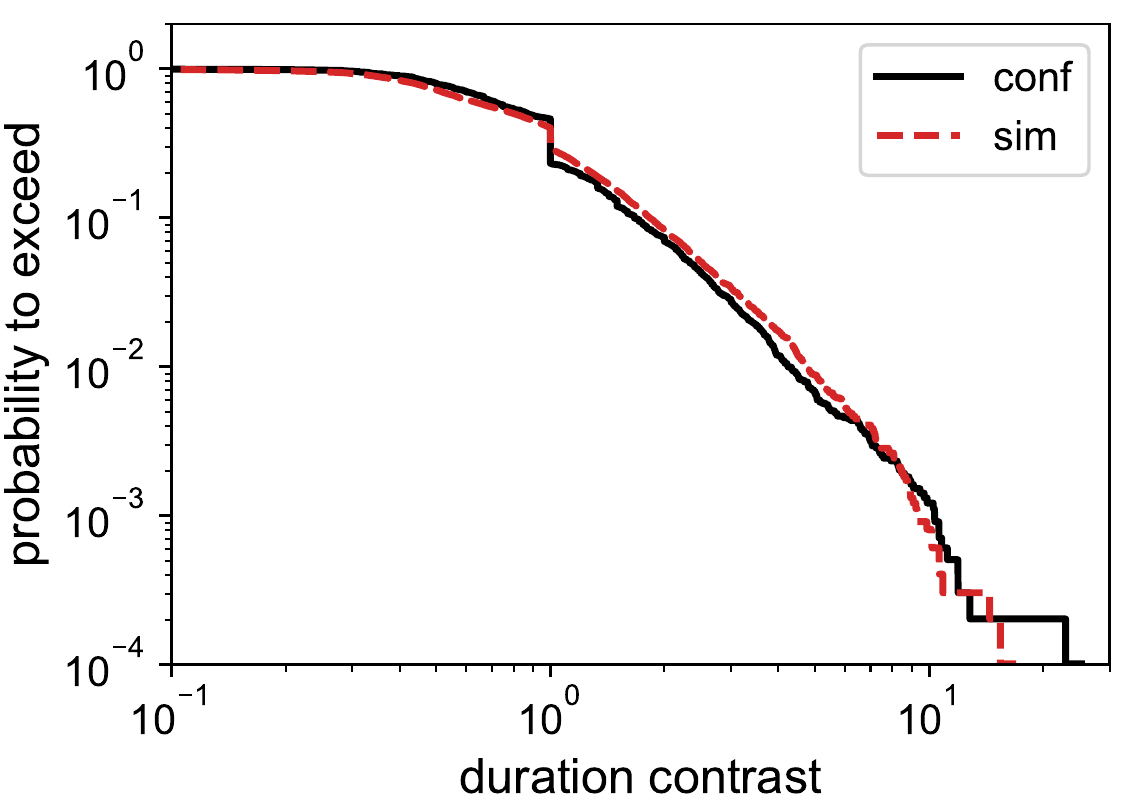}}
  \subfigure[\malawi]{\includegraphics[width=0.49\textwidth]{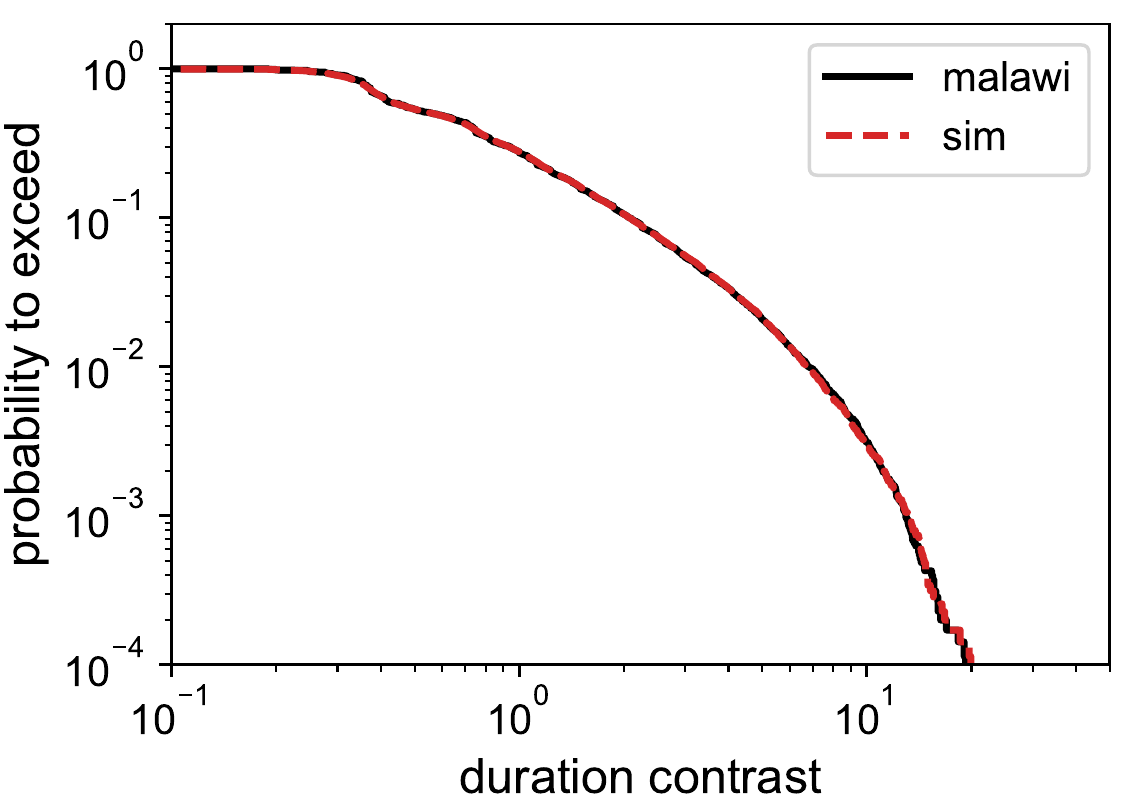}}
  \subfigure[\baboons]{\includegraphics[width=0.49\textwidth]{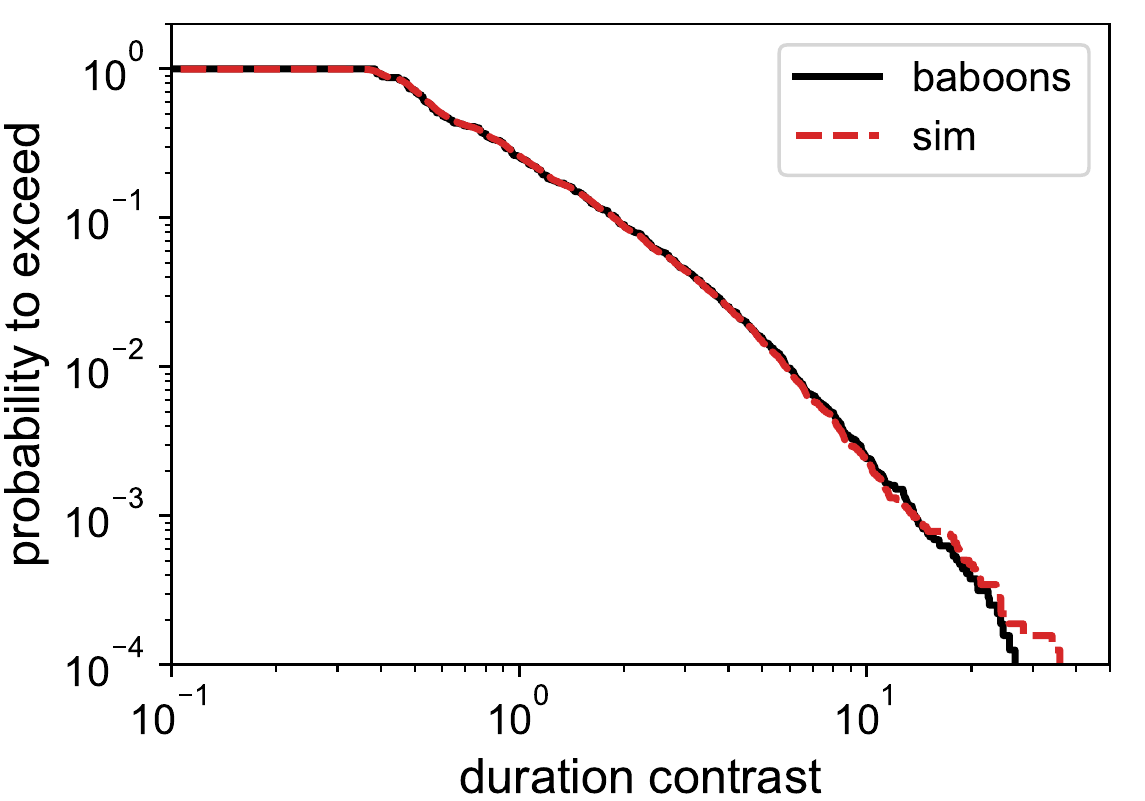}}
\caption{\label{fig:mc} Results of the simulations described in the
  manuscript (Sect 3) without using a cut on \Ncon for all the datasets.}
\end{figure}

We have shown in the manuscript how to reproduce the noise present on
the data for all $\Nint$ values on the \conf dataset in Section 3.
\Fig{mc} shows the results for the other datasets. We
reproduce all the data, using only a small
fraction of the timelines (the ones with $\Ncon>50$ which represents
respectively ).

\subsection{Contrast per relation}

In the previous section, the duration (and contrast) of all contacts are combined on
\Fig{mc} in the sense the contrast of each relation are mixed together.
We now consider each relation and show its contrast \pte with
different colors. The timelines are noisy since we do not use any
minimal number of steps (\Nint). The \baboons result is less noisy but this
is only due to the fact that the timelines have more samples due to
the longer data-taking period (26 days).

\begin{figure}[ht!]
  \centering
  \subfigure[\hosp]{\includegraphics[width=0.49\textwidth]{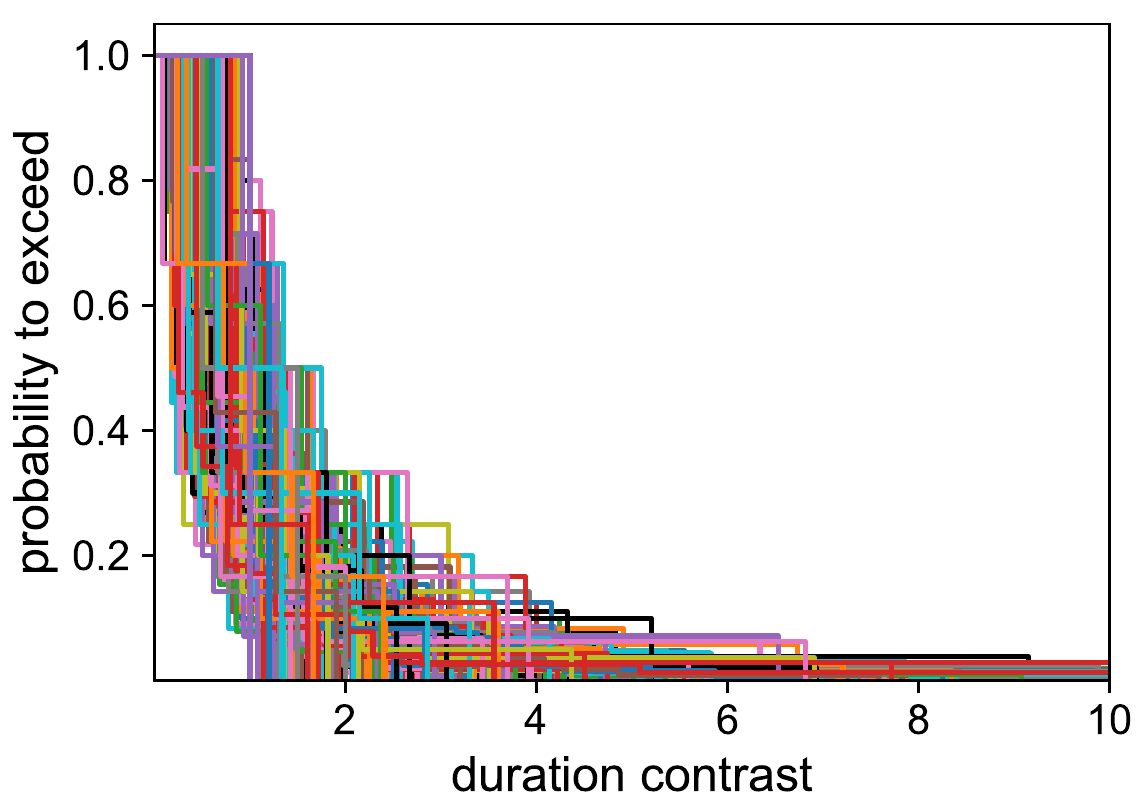}}
  \subfigure[\conf]{\includegraphics[width=0.49\textwidth]{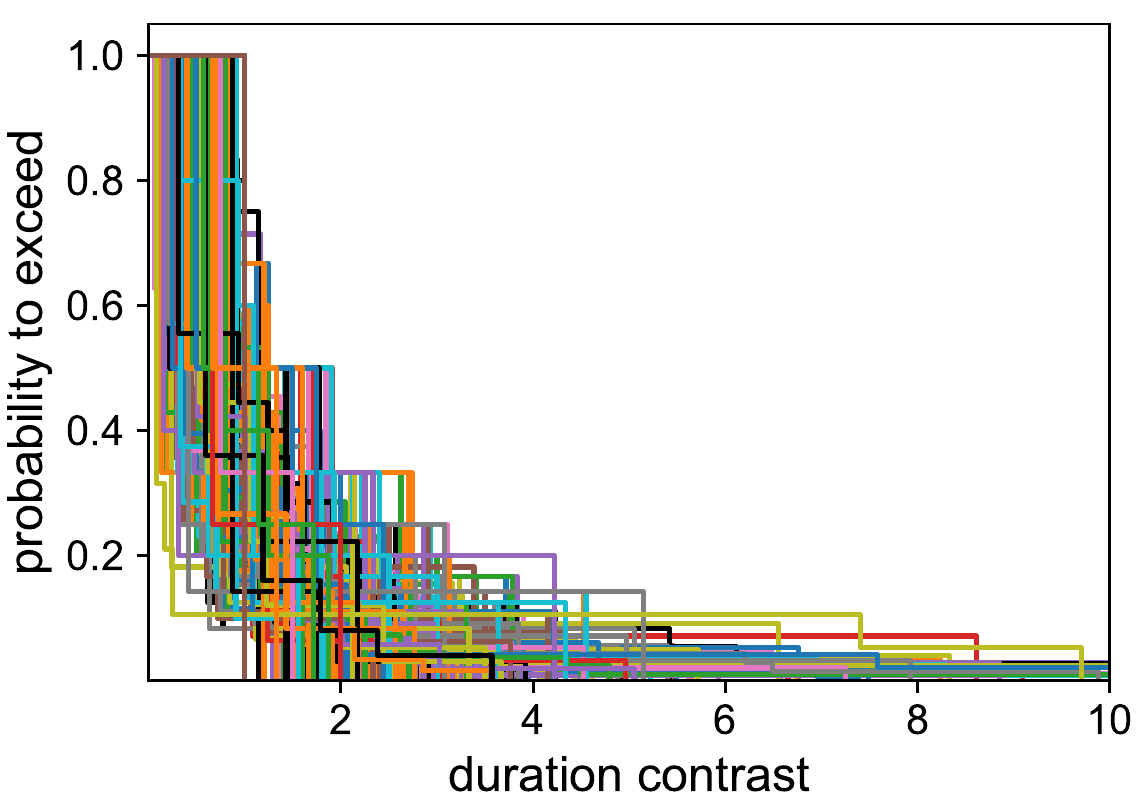}}
  \subfigure[\malawi]{\includegraphics[width=0.49\textwidth]{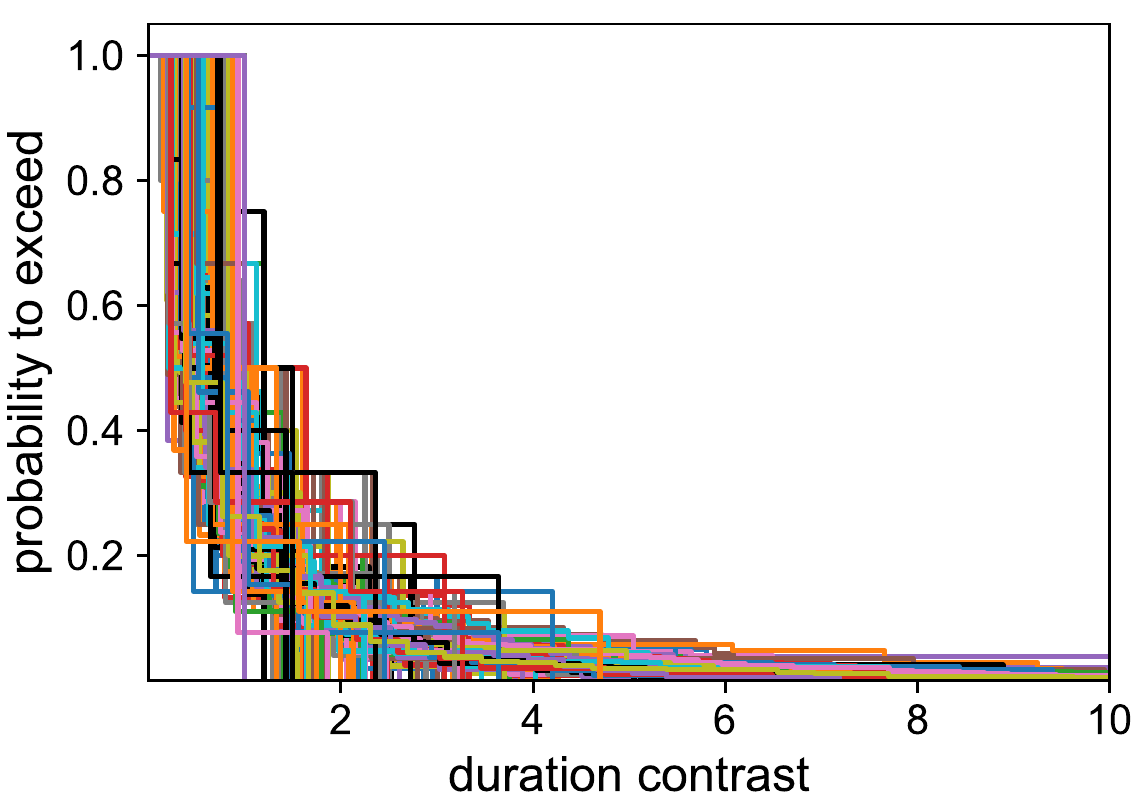}}
  \subfigure[\baboons]{\includegraphics[width=0.49\textwidth]{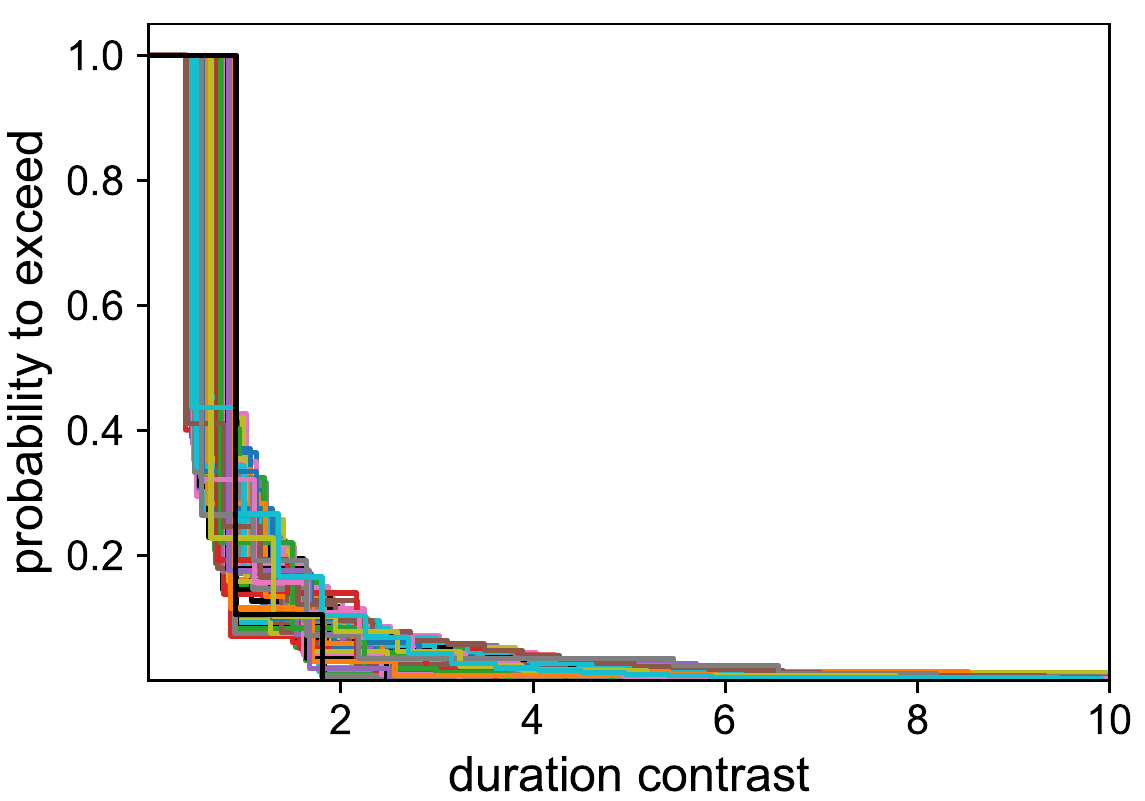}}
\caption{\label{fig:cdc_all} Contrast distributions per relation
  measured on the data without any \Nint cut. Each color represents a different
  relation timeline.}
\end{figure}

\begin{figure}[ht!]
  \centering
  \subfigure[\hosp]{\includegraphics[width=0.49\textwidth]{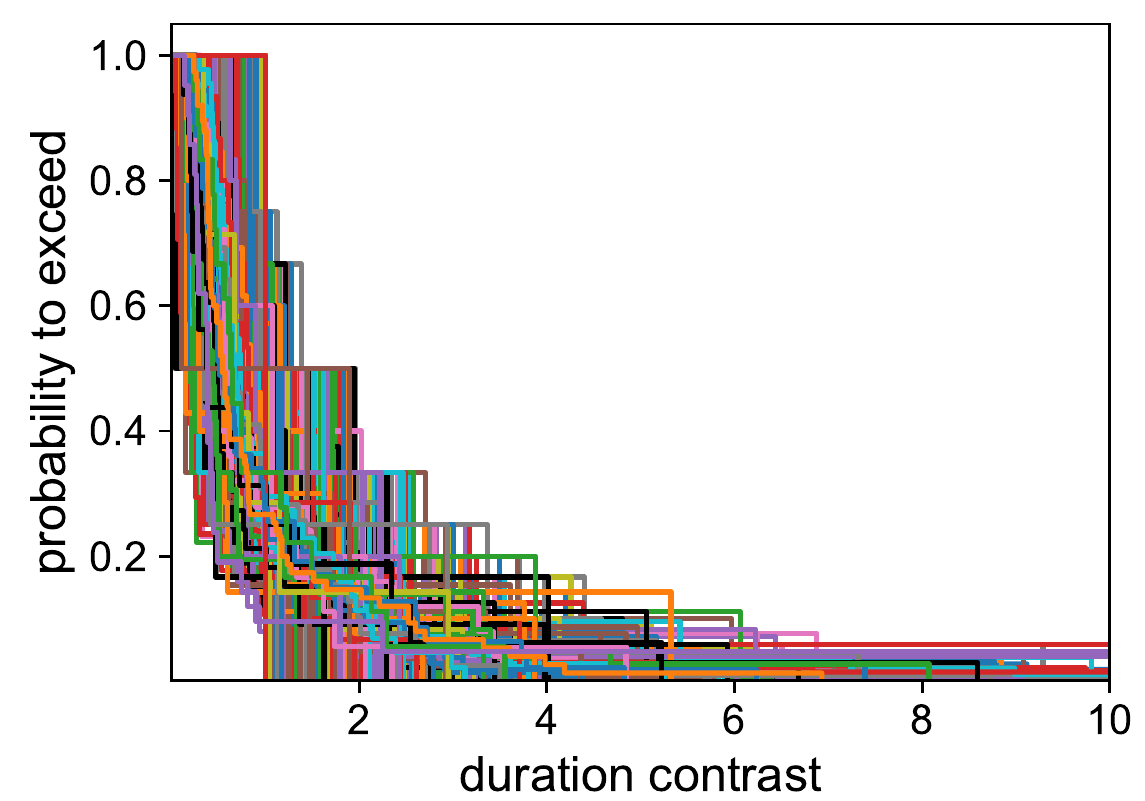}}
  \subfigure[\conf]{\includegraphics[width=0.49\textwidth]{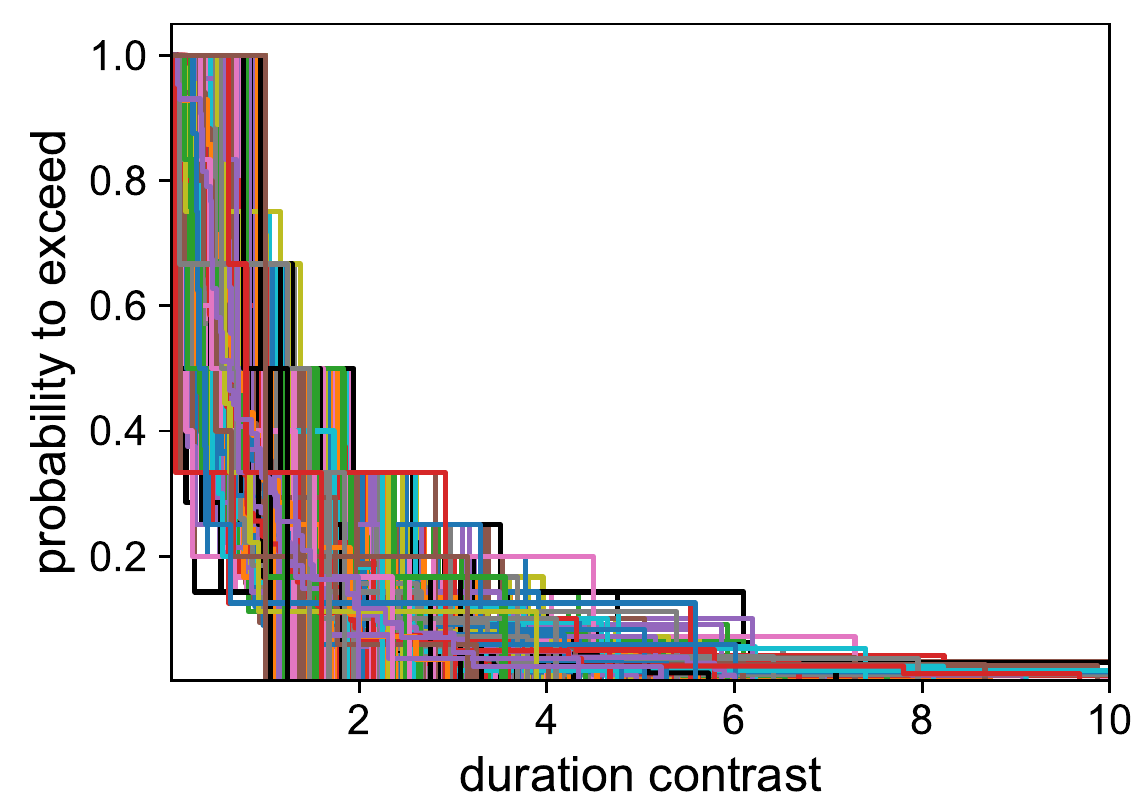}}
  \subfigure[\malawi]{\includegraphics[width=0.49\textwidth]{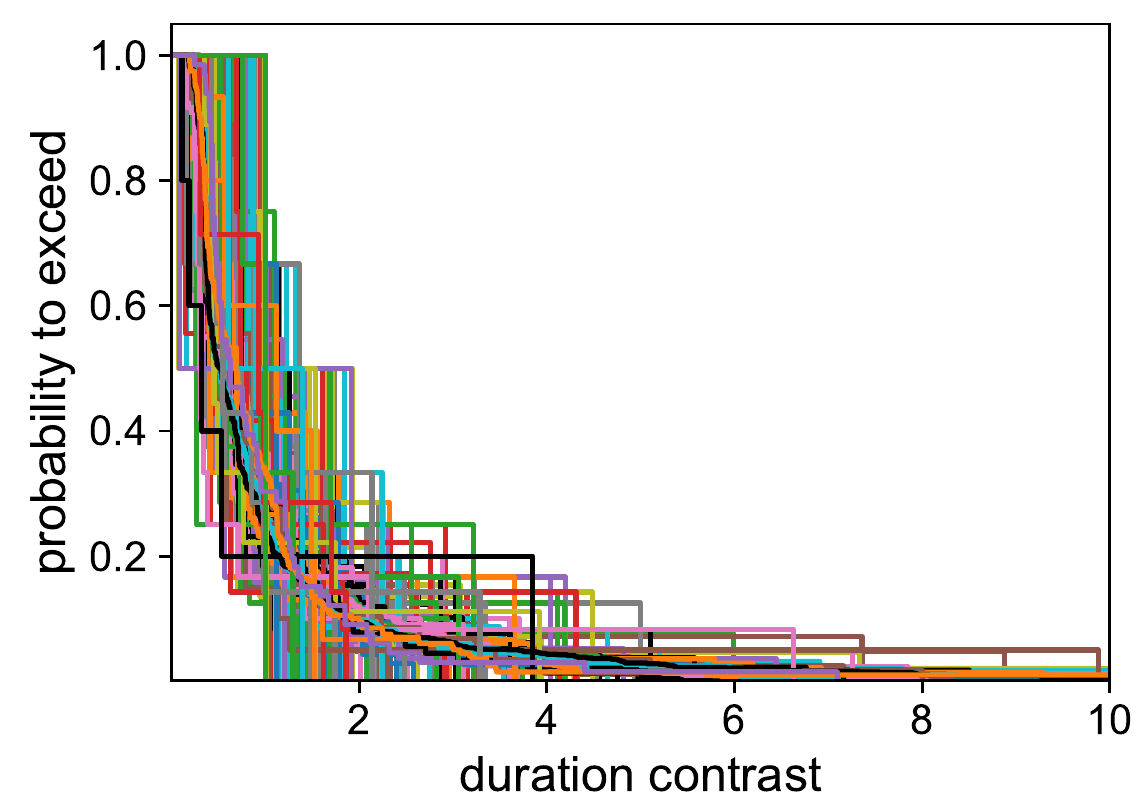}}
  \subfigure[\baboons]{\includegraphics[width=0.49\textwidth]{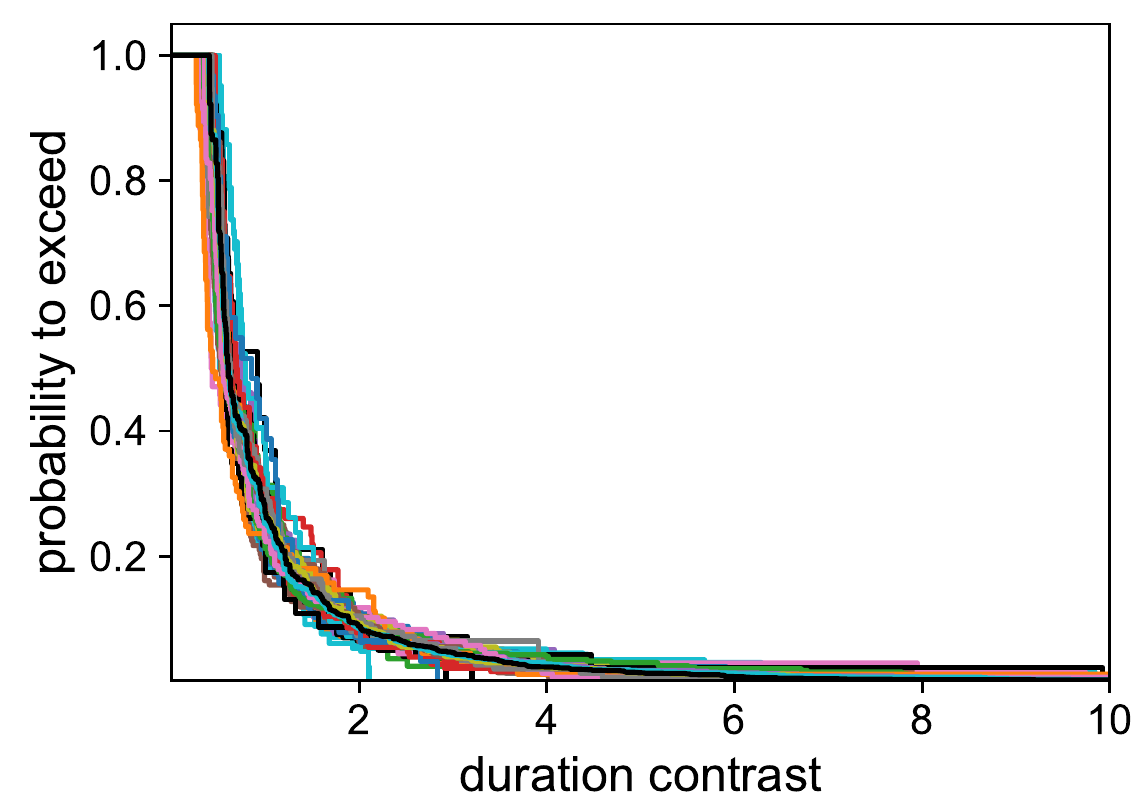}}
\caption{\label{fig:sim_all} Results of the simulations described in the
  manuscript (Sect 3) for each relation (colored curves) 
without using a cut on \Ncon for all the datasets.}
\end{figure}

We then perform the same simulations than described in the manuscript, but this
time on each relation individually. Once again only the global shape
of the contrast obtained with the $\Ncon>50$ cut (i.e. \Fig{cdc}(b)) is used to
draw the random numbers. The results are shown on \Fig{sim_all}. They
resembles closely the data (\Fig{cdc_all}). This confirms 
that the shape and spread observed on data (\Fig{cdc_all})
can be reproduced using only the cleaned \Fig{cdc} distribution
($\Ncon(r)>50$ and statistical fluctuations (on the mean) due to
the limited size statistics.

\section{Number of clusters in Levy graphs}
\label{sec:A}

As shown in \cite{Plaszczynski:2022}, the mean fraction of clusters
follows a power-law function of the scale with a relatively small
spread among realizations
\begin{align}
\label{eq:Nclus}
  \dfrac{\Nclus}{N}=\dfrac{A}{s^{\alpha_c}}
\end{align}

\begin{figure}[ht!]
  \centering
  \includegraphics[width=0.7\textwidth]{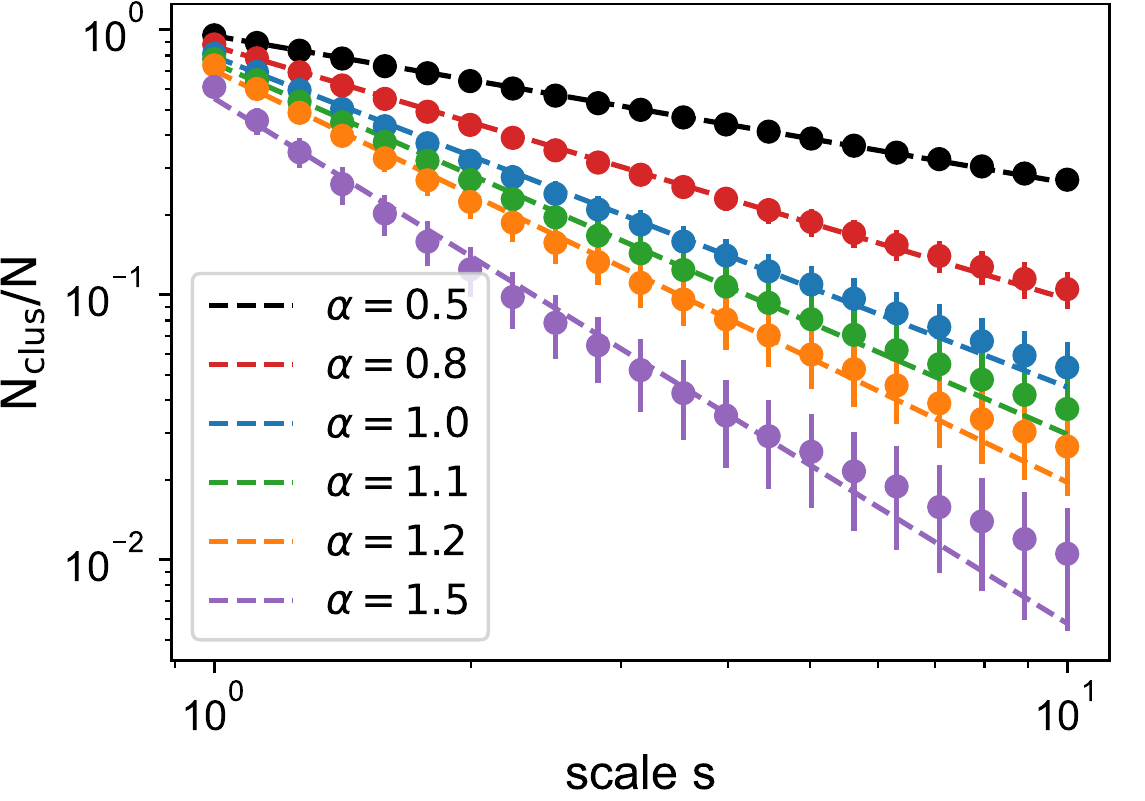}
\caption{\label{fig:fit}Mean fraction of clusters as a function of the
scale as determined from LGG simulations for several $\alpha$ values.
Dashed lines represent the best power-law fits.}
\end{figure}

$A$ and $\alpha_c$ are determined using 1000 realizations of LGGs for
a fixed $\alpha$ value and varying the scale $s$ between 1 and 10.
\Fig{fit} shows the results and the power-law fits. The resulting
parameters are given in Table \ref{tab:fit}

\begin{table}
  \centering
\begin{tabular}{rcc}
\toprule
  $\alpha$ &    A &   $\alpha_c$ \\
\midrule
0.50 & $0.95 \pm0.01$ & $0.56\pm0.01$ \\
0.80 & $0.87\pm0.02$ & $0.95\pm0.02$ \\
1.00 & $0.79\pm0.02$ & $1.25\pm0.04$ \\
1.10 & $0.75\pm0.02$ & $1.40\pm0.05$ \\
1.20 & $0.70\pm0.03$ & $1.55\pm0.06$ \\
1.50 & $0.55\pm0.03$ & $1.98\pm0.09$ \\
\bottomrule
\end{tabular}
\caption{\label{tab:fit} Coefficients of \refeq{Nclus} fitted 
  from the 2D LGG simulations (\Fig{fit}) for various $\alpha$ values.}
\end{table}

\section{Poisson duration}
\begin{figure}[ht!]
  \centering
  \includegraphics[width=0.7\textwidth]{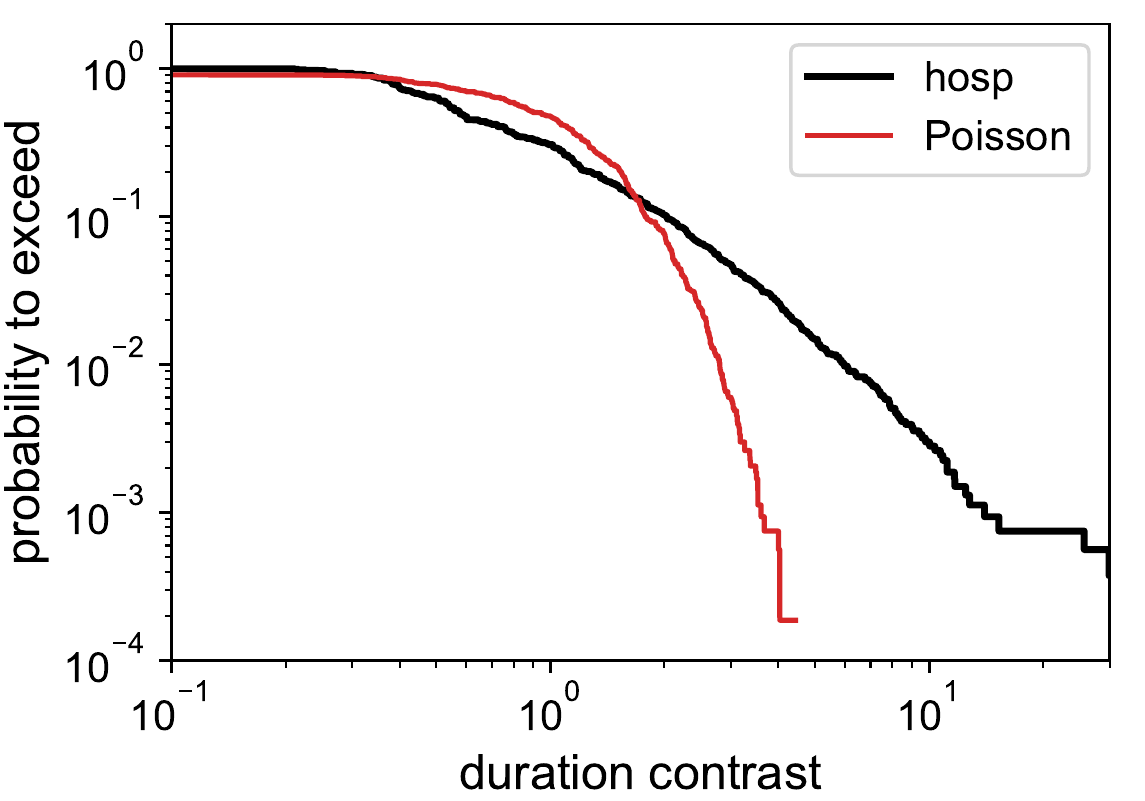}
\caption{\label{fig:Poisson} Distribution of the contrast durations
  assuming Poisson distributed durations. The
  parameters are taken from the \hosp
  dataset and the data result is recalled in red.}
\end{figure}

Dividing my mean-values does not
necessarily lead to distributions of the \Fig{cdc} kind.
We illustrate that feature by computing the contrast using
Poisson-distributed durations. To this purpose we use the \hosp data
to obtain the $\bar t(r)$ values for each relation; we then draw $\Ncon(r)$
random numbers following a Poisson distribution of parameter $\bar
t(r)$, determine the arithmetic mean and compute the contrast. The result is show on \Fig{Poisson}
which is clearly different from the results observed on data.

\section{Dimension above 2}

\begin{figure}[ht!]
  \centering
  \includegraphics[width=0.7\textwidth]{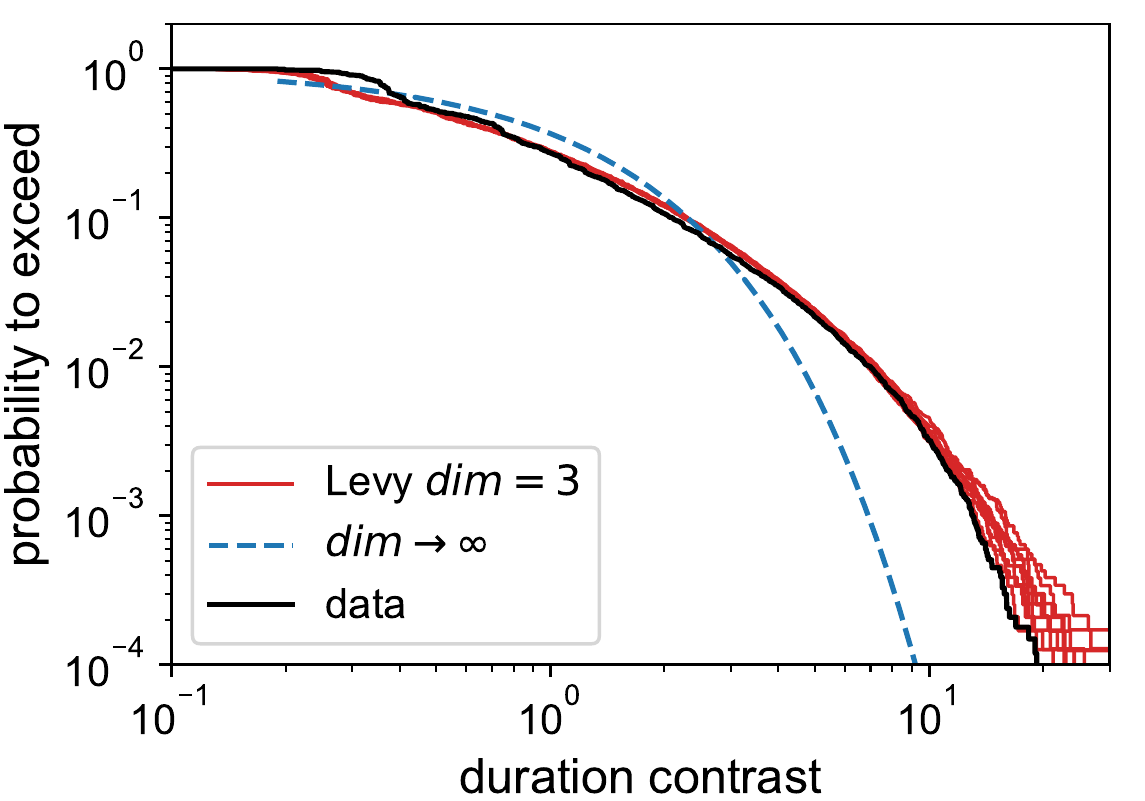}
\caption{\label{fig:dim3} Levy model ($\alpha=1.1$) in red compared to
the \malawi data (in black) for a Levy walk in a space of dimension 3.
Increasing the dimension further, the model converges to the dashed
blue line. }
\end{figure}

A Levy flight can be generalized to any dimension (see
\cite{Plaszczynski:2022} for a simple algorithm). For our model, we
have tried several dimension. \Fig{dim3} shows the result in dimension
3 (note that the coefficients  $A$ and $\alpha_c$ from \sect{A} have
been recomputed). Compared to dimension 2 (see \Fig{models} (c)), the
agreement with the \malawi data gets worse. If we  further increase
the dimension the return probability of the walk decreases and
the contrast goes to $\exp(-\delta)$ which is clearly far from the data.

\section{Effect of the time resolution}

The timing resolution of the experiments affects our model in two
ways. The size and scale of each LGG (thus relation) are given by
\begin{align}
  N(r)&=\sum_{i=1}^{\Nint(r)} t_i(r) \\
  s(r)&=\left[A \bar t(r)\right]^{1/\alpha_c}
\end{align}
where $A\simeq 1, \alpha_c\simeq \alpha$ (see \sect{A}).

\begin{figure}[ht!]
  \centering
  \includegraphics[width=0.7\textwidth]{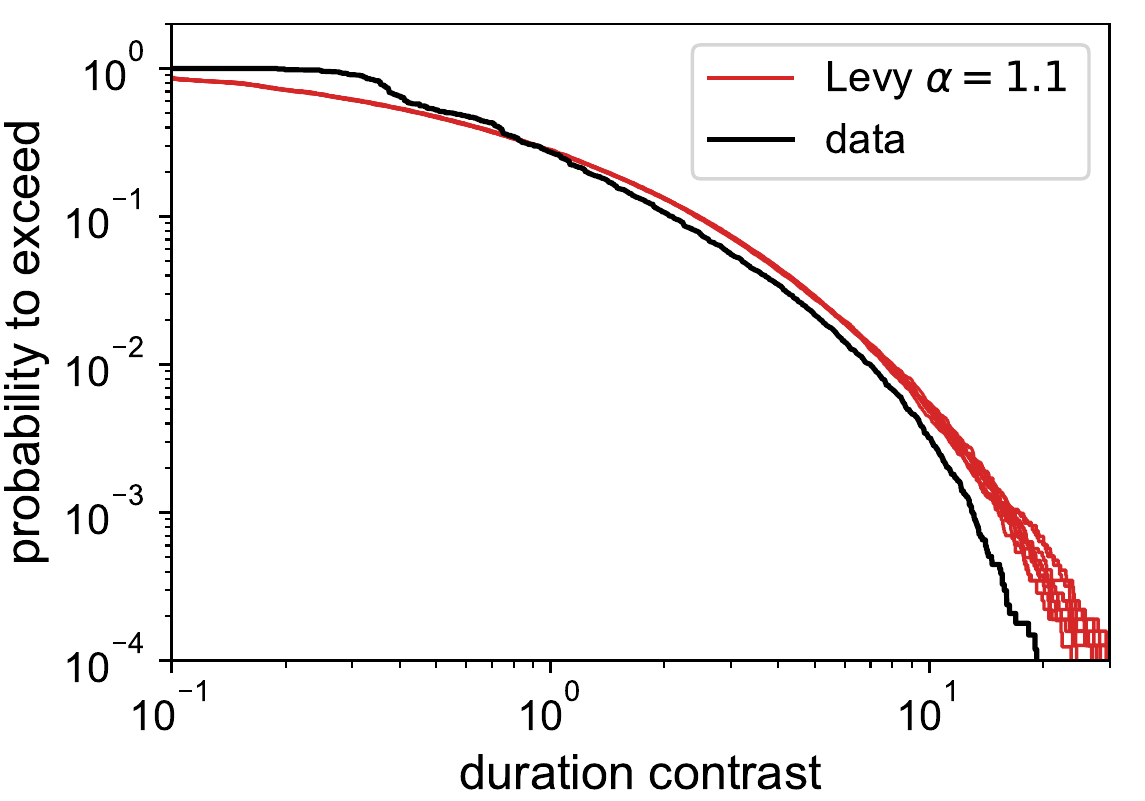}
\caption{\label{fig:Tresol} Levy model ($\alpha=1.1$) modified 
by increasing the resolution to T=5 s (red curves) compared to the
\malawi dataset.}
\end{figure}

$t_i$ and $\bar t$ are expressed as a number of resolution steps (which in
the paper is $T=20$ s). We may artificially change it, assuming for
instance a resolution of $T=5$ s, in which case we multiply the $t_i$'s and
$\bar t$ by a factor 4. This changes for each graph the
values of $N$ and $s$. The result on the contrast distribution for
this model are shown on \Fig{Tresol}.

Compared to \Fig{models}(c) this model fits less well the data
which is normal since we have used a fake resolution. It shows that
the wiggles at the beginning of the distribution are due to the $T=20$
s resolution. A prediction from our model is thus that the contrast
distribution should be smoother and slightly different if we have data
with a better timing resolution.

\newpage
\bibliographystyle{unsrt}
\bibliography{references}